\documentclass[fleqn,usenatbib]{mnras}

\usepackage{newtxtext,newtxmath}

\usepackage[T1]{fontenc}
\usepackage{ae,aecompl}
\usepackage{rotating}
\usepackage{color}

\usepackage{graphicx}	
\usepackage[utf8]{inputenc}
\usepackage[dvipsnames]{xcolor}
\definecolor{royalfuchsia}{rgb}{0.79, 0.17, 0.57}

\title{Comprehensive X-ray and Multi-wavelength Study of ULXs in NGC 1566}

\author[S. Allak]{Sinan Allak\thanks{E-mail:0417allaksinan@gmail.com}
\\
Space Science and Solar Energy Research and Application Center (UZAYMER), University of Çukurova, 01330, Adana, Turkey\\
}

\date{Accepted 2023 November 24. Received 2023 November 17; in original form 2023 July 28
}

\pubyear{2023}

\begin{document}
\label{firstpage}
\pagerange{\pageref{firstpage}--\pageref{lastpage}}
\maketitle

\begin{abstract}
This paper presents a comprehensive X-ray and multi-wavelength study of ultraluminous X-ray sources (ULXs) in NGC 1566 using archival {\it Chandra}, {\it Swift/XRT}, {\it James Webb Space Telescope, JWST}, and {\it Hubble Space Telescope, HST} observations. The main results are, first, from the hardness ratio diagram, where spectral state transitions from hard to soft as seen in typical Galactic high mass X-ray binaries for ULX-3 was observed. Second, a new transient ULX candidate (ULX-4) was identified, reaching a peak luminosity of $\sim$10$^{40}$ \textit{erg} s$^{-1}$. Third, the optical and NIR (near-infrared) counterparts of the ULXs were searched from the precise astrometric calculations. For ULX-1 and ULX-2, evidence was found that the observed NIR emission is due to the circumbinary disk/dust disrupted by X-rays. Fourth, the optical observations suggest that the possible donor star of ULX-3 is a B-type supergiant. In the case of ULX-4, the multi-wavelength properties are not clear since many sources are detected within the astrometric error radius.

\end{abstract}


\section{Introduction} 
\label{sect:intro}

Ultraluminous X-ray sources (ULXs) are a sub-class of X-ray binaries (XRBs) with an isotropic X-ray luminosity (L$_{X}$ $\geq$ 10$^{39}$ erg s$^{-1}$)exceeding the Eddington limit of a 10 M$\odot$ black hole (BH). Almost a consensus has been reached that ULXs are thought to be powered by supercritical accretion onto neutron stars or stellar mass black holes (sMBHs) (see review by \citealp{2017ARA&A..55..303K,2021AstBu..76....6F,2023NewAR..9601672K} and \citealp{2023arXiv230200006P}). On the other hand, many studies suggested the existence of intermediate-mass black holes (IMBHs, 10$^{2}$- 10$^{5}$ M$\odot$) accreting at sub-Eddington rates for the compact object nature of ULX systems \citep{2007Ap&SS.311..203R,2012MNRAS.423.1154S,2013MNRAS.436.3262C,2013ApJ...774L..16P}. Alternatively, geometric beaming and/or accretion at super-Eddington rates on stellar mass compact objects can explain the observed properties of ULXs \citep{2007Ap&SS.311..203R,2009MNRAS.393L..41K,2015MNRAS.454.3134M,2018ApJ...857L...3W}.\\

Temporal variability is a crucial aspect observed for ULXs and accretion-powered sources. The variability properties of the many Galactic XRBs and ULXs have unveiled quasi-periodic oscillations \citep{2015MNRAS.446.3926A}, and coherent pulsations (e.g., \citealp{2014Natur.514..202B,2018MNRAS.476L..45C,2020ApJ...895...60R}) that can provide insights into the nature of the compact object. Moreover, some study of the long-term variability of ULXs has shown orbital or super-orbital variations (e.g., \citealp{2019ApJ...873..115B}). Thanks to {\it Neil Gehrels Swift Observatory X-ray Telescope} (hereafter {\it Swift/XRT}) it is possible to obtain long and regularly sampled light curves of ULXs \citep{2015A&A...580A..71L,2020MNRAS.491.1260S,2022ApJ...929..138B,2022MNRAS.510.4355A}. Therefore, comparing the variability properties of Galactic XRBs, which are well-known, with ULXs gives us the chance to place strong constraints on compact objects. Moreover, the discovery of a cyclotron resonance scattering feature (CRSF), which is directly related to the measurement of the magnetic field, in the X-ray spectrum of ULXs indicates the presence of a neutron star (NS) candidate \citep{2018NatAs...2..312B, 2022ApJ...933L...3K}.\\

Studying Galactic XRBs enhances our understanding of the characteristics of the donor star of ULXs, the mass of the compact object, the existence of focused outflows known as jets, and the surrounding environment \citep{2023arXiv230200006P}. The observed optical emission in ULXs can originate from the accretion disk, the donor star, or a combination of both \citep{2011ApJ...737...81T,2022MNRAS.515.3632A,2022MNRAS.517.3495A}. In the many studies ( e.g., \citealp{2012MNRAS.420.3599S,2018MNRAS.480.4918A,2019ApJ...884L...3Y,2014MNRAS.444.2415S}), focusing on optical variability, multi-band colors, and SED modeling, strongly suggest that the optical emission is likely contaminated or even dominated by reprocessed radiation from an irradiated accretion disk. Additionally, the super-orbital or orbital period (sinusoidal modulations) has been determined for a few optical counterparts in long-term optical light curves \citep{2022MNRAS.510.4355A,2022MNRAS.517.3495A,2009ApJ...690L..39L}.\\

Such as dynamical mass construction, considerable study has been conducted to determine the donor stars of ULXs \citep{2012ApJ...745...89L}. However, this is quite difficult to do for all ULXs because their apparent magnitudes (m$_{V}$) are usually faint (m$_{V}$ $\leq$ 19 mag). Moreover, they are located in star-forming and crowded fields, therefore, space detectors with good enough spatial resolution are needed. The significant improvement in sensitivity and resolution provided by the {\it JWST} may allow us to identify counterparts in such fields with a high degree of confidence. The studies of \cite{2023arXiv230611163A,2023MNRAS.526.5765A} showed that even in the {\it HST} images, some optical counterparts appear to be uniquely blended sources. Previous studies (e.g., \cite{2014MNRAS.442.1054H,2020MNRAS.497..917L}) have claimed that some counterparts of ULXs, bright in the NIR band, might be red supergiants (RSGs). Since ULXs are located in crowded regions of the host galaxies, the past images of many of the point-like and/or bright NIR counterparts are likely unresolved sources in the distant galaxies, therefore, many sources may not be red color enough for them to be an RSG 
 \citep{2023arXiv230611163A,2023MNRAS.526.5765A}.\\

Cold astronomical objects that are not detectable by optical detectors can be observed in the infrared (IR) bands. ULXs whose counterparts could not be determined, in other words, no optical source could be determined in the derived astrometric error radii in the past studies, are highly probable to have counterparts in the good enough resolution {\it JWST} images. Recently, in some studies, XRBs including ULXs whose optical counterparts cannot be detected have been classified as low-mass X-ray binaries (LMXBs) \citep{2020ApJ...890..150C}. However, recently published studies \citep{2023arXiv230611163A,2023MNRAS.526.5765A} have shown evidence of circumbinary disk/dust around ULXs whose optical counterparts could not be determined. These results show that the emission of optical counterparts (possible donor stars) can be obscured by hot dust and/or circumbinary disks around them, which does not imply that they are LMXBs. \\

This work carries out a comprehensive X-ray and multi-wavelength study of the ULXs in the galaxy NGC 1566 observed with {\it Swift/XRT}, {\it Chandra}, {\it JWST} and {\it HST}. The primary goal of this work is to search for and identify possible NIR and optical counterparts of ULXs by deploying all available {\it JWST} and {\it HST} images. In addition, the X-ray spectral and temporal properties of these ULXs are analyzed in detail, aiming to impose constraints on the nature of the compact objects nature as well as on their X-ray emission. The paper is structured as follows. Section \ref{sec:2} presents the properties of galaxy NGC 1566 as well as the target ULXs, and also it gives details of X-ray, optical, and infrared observations. Details of these observations and data reduction and analysis are presented in Section \ref{sec:3}. Section \ref{sec:4} presents results and discussions of the properties of X-ray and multi-wavelength of ULXs. Finally, Section \ref{sec:5} summarizes the main conclusions of this study.\\

\section{Target of ULXs \& Observations}
\label{sec:2}

\begin{figure*}
\begin{center}
\includegraphics[angle=0,scale=0.27]{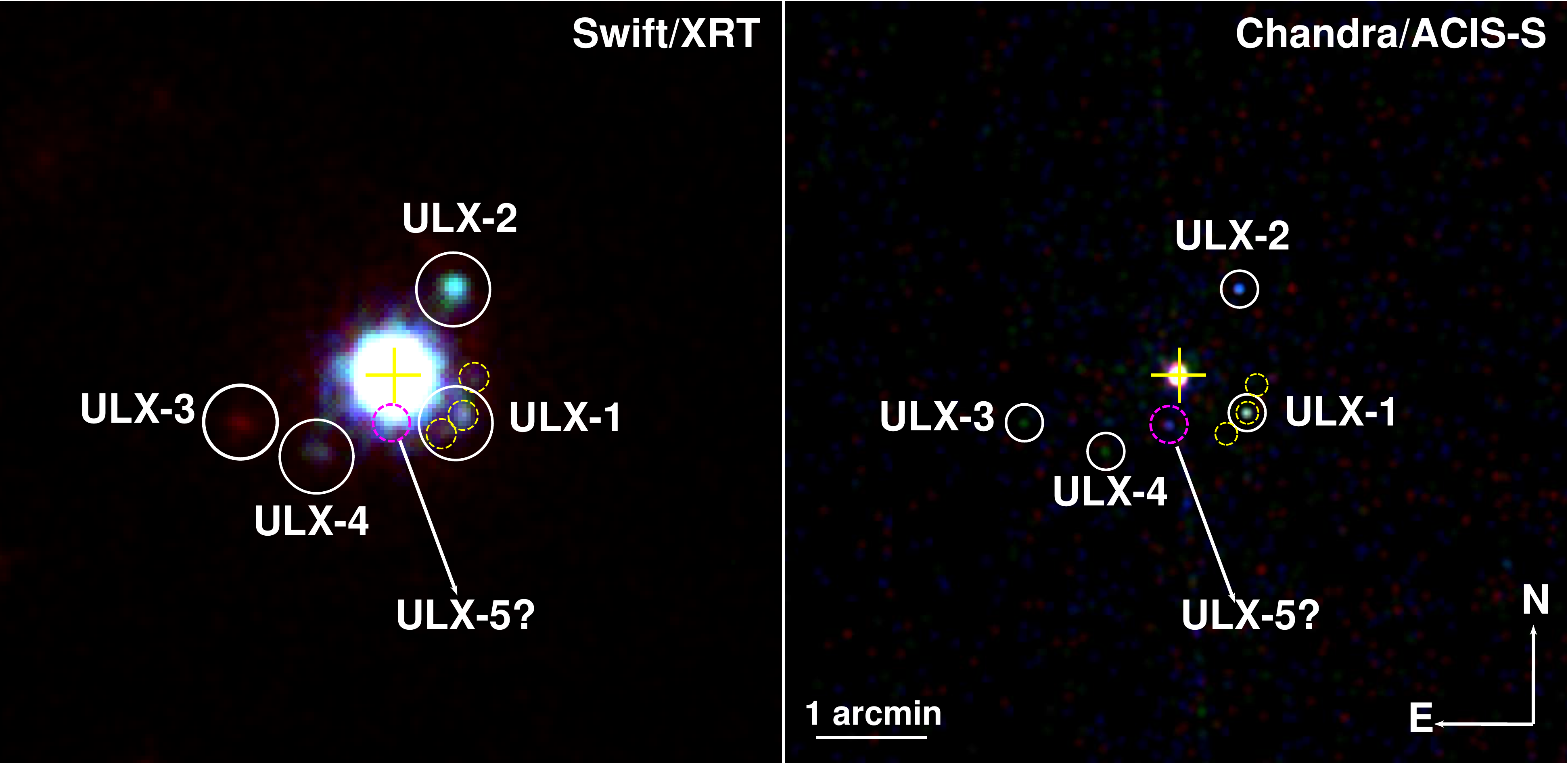}
\caption{The locations ULXs in NGC 1566 on the true color (Red: 0.3-1 keV; Green: 1-2 keV; Blue: 2-8 keV) merged image of all {\it Swift/XRT} observations (left) and {\it Chandra} (right). The images are smoothed with a 3 arcsec Gaussian. The ULXs are indicated with white circles, and dashed yellow circles indicate sources detectable on {\it Swift/XRT} at around ULX-1. Both images are on the same scale.}
\label{F:RGB}
\end{center}
\end{figure*}

\begin{figure*}
\begin{center}
\includegraphics[angle=0,scale=0.27]{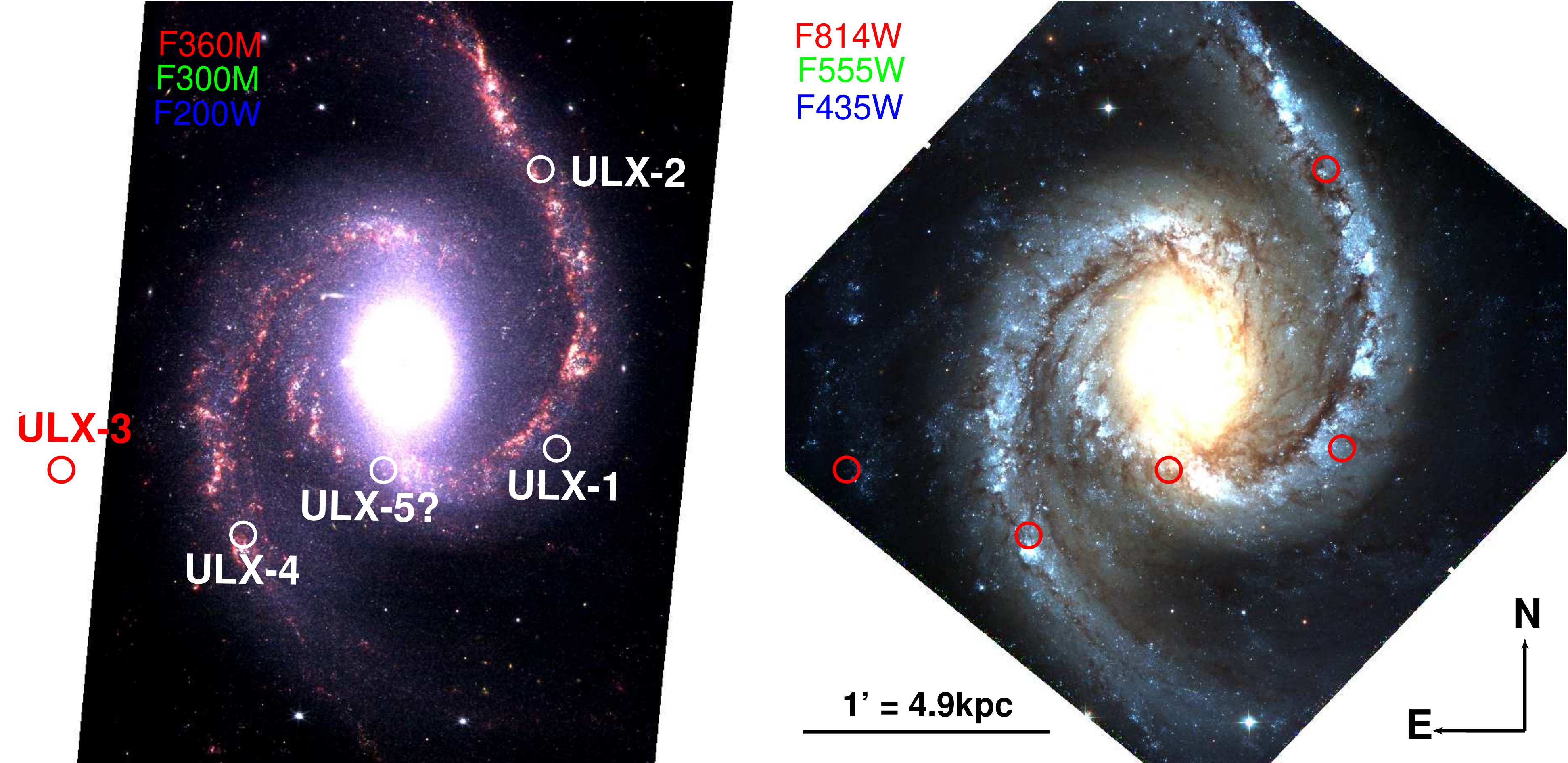}
\caption{The RGB (Red;Green;Blue) images {\it JWST} (left) and {\it HST} (right) of the galaxy NGC 1566. The filters used for RGB images and positions of ULXs are shown on the images. The images are smoothed with 3 arcsec Gaussian. Both images are on the same scale.}
\label{F2}
\end{center}
\end{figure*}

\cite{2005ApJS..157...59L} presented several ULX samples in nearby galaxies through observations made with the\textit{ ROSAT High-Resolution Imager (ROSAT/HRI)}. Among these samples, they identified three ULXs located in the face-on barred Sbc spiral galaxy NGC 1566. In their study, the X-ray luminosity ranges of these three ULXs were given as L$_{X}$ $\sim$ (1.36-6.5) $\times$ 10$^{39}$ erg s$^{-1}$, and these ULXs showed variability of approximately a factor of 6 in {\it ROSAT/HRI} observations based on a distance of 13.4 Mpc. The locations of these sources are shown in Figure \ref{F:RGB}. As seen in Figure \ref{F2}, ULX-1 and ULX-2 are located in the thin spiral arms, and ULX-3 is located on the edge of a spiral arm. Furthermore, ULX-3 was not observed in the {\it JSWT} observations. The host galaxy NGC 1566 of ULXs is a massive, face-on star-forming (SFR = 4.5 M$\odot$ yr$^{-1}$) spiral galaxy at a distance of 17.69 Mpc (\citealp{2023arXiv230105718M} and references therein). This distance is used throughout this study.\\

The galaxy NGC 1566 was observed on November 22, 2022, by using both the {\it JWST} near-infrared camera instrument ({\it NIRCam}) and mid-infrared instrument ({\it MIRI}) (GO program 2107, PI: J. Lee). The {\it JWST} observations include {\it NIRCam} imaging by using the F200W, F300M, F335M, and F360M filters and {\it MIRI} imaging using the F770W, F1000W, F1130W, and F2100W filters. NGC 1566 was also observed by {\it HST} on September 03, 2013, using WFC3/UVIS (The Wide Field Camera 3) (GO program 13364). The {\it HST} observations include UVIS imaging by using F275W, F336W, F438W, F555W, and F814W filters. Moreover, NGC 1566 was also observed during 3 ks by {\it Chandra} ACIS (Advanced CCD Imaging Spectrometer) in 2006 (Obs ID: 21478), and it was observed many times by {\it Swift/XRT} between 2011 and 2022. Additionally, the {\it GAIA} DR3\footnote{https://www.cosmos.esa.int/web/gaia/data-release-3} source catalog is used for astrometric calculations. The details of observations are given in Table \ref{T1}.

\begin{table*}
\centering
\caption{The Log of observations of the galaxy NGC 1566}
\begin{tabular}{ccccccccccccccccllll}
\hline
\\
Observatory/Instrument & Proposal ID & Date & Exp & Filter \\
& & (YYYY-MM-DD) & (s)\\
\hline
HST/WFC3/UVIS & 13364 & 2013-09-03 & 2382 & F275W \\
HST/WFC3/UVIS & 13364 & 2013-09-03 & 1119 & F336W \\
HST/WFC3/UVIS & 13364 & 2013-09-03 & 965 & F438W \\
HST/WFC3/UVIS & 13364 & 2013-09-03 & 1143 & F555W \\
HST/WFC3/UVIS & 13364 & 2013-09-03 & 989 & F814W \\ 
HST/WFC3/IR & 15133 & 2018-04-06 & 596.9 & F160W \\ 
\\
JWST/MIRI & 2107 & 2022-11-22 & 532.8 & F770W \\
JWST/MIRI & 2107 & 2022-11-22 & 732.6 & F1000W \\
JWST/MIRI & 2107 & 2022-11-22 & 1864.8 & F1130W \\
JWST/MIRI & 2107 & 2022-11-22 & 1931.4 & F2100W \\
JWST/{\it NIRCam} & 2107 & 2022-11-22 & 2405.0 & F200W \\
JWST/{\it NIRCam} & 2107 & 2022-11-22 & 773.1 & F300M \\
JWST/{\it NIRCam} & 2107 & 2022-11-22 & 773.1 & F335M \\
JWST/{\it NIRCam} & 2107 & 2022-11-22 & 858.9 & F360M \\
\hline
\\
& Target ID & Start time & N$^{a}$ & Exp$^{b}$ \\
& & (YYYY-MM-DD $-$ YYYY-MM-DD)& & (ks)& \\
\hline
Swift/XRT & 00014916 & 2021-11-11 $-$ 2022-01-17 & 27 & 55.53 \\
Swift/XRT & 00014923 & 2021-11-18 $-$ 2021-11-19 & 2 & 8.67 \\
Swift/XRT & 00015302 & 2022-08-23 $-$ 2022-08-26 & 4 & 7.99 \\
Swift/XRT & 00031742 & 2010-06-23 $-$ 2020-10-21 & 13 & 15.66 \\
Swift/XRT & 00033411 & 2014-09-11 $-$ 2015-07-30 & 28 & 52.10 \\
Swift/XRT & 00035880 & 2007-12-12 $-$ 2019-07-31 & 133 & 123.21 \\
Swift/XRT & 00045604 & 2011-08-25 $-$ 2020-07-20 & 43 & 44.22 \\
Swift/XRT & 00088910 & 2019-08-08 $-$ 2019-08-21 & 3 & 5.55 \\
Swift/XRT & 03111666 & 2022-06-21 $-$ 2022-08-19 & 9 & 5.30 \\
\hline
\\
& Obs ID & Date & Exp \\
& & (YYYY-MM-DD) & (ks)& \\
\hline
Chandra/ACIS-S	& 21478 & 2018-12-03 & 3.00 \\
\hline
Notes: $^{a}$ Number of observations $^{b}$ Total exposure time\\
\end{tabular}
\label{T1}
\end{table*}

\section{Data Reduction and Analysis} \label{sec:3}

\subsection{Chandra and Swift/XRT}

{\it Chandra} ACIS-S observation was analyzed by using {\it Chandra} Interactive Analysis of Observations ({\scshape ciao})\footnote{https://cxc.cfa.harvard.edu/ciao/} v4.15 software and calibration files {\scshape caldb} v4.10. The level 2 event files were obtained with {\it chandra\_repro} in {\scshape ciao}. The {\it wavdetect} tool in {\scshape ciao} was used for source detection, and eight X-ray sources were identified in \textit{Chandra} image, including the three ULXs previously identified by \cite{2005ApJS..157...59L}. The source and the background events were extracted from circular regions of 4 and 8 arcsec radius. The source spectra and light curves were obtained with the tasks {\it specextract} and {\it dmextract}, respectively. In the case of \textit{Chandra} spectra, by using the task {\it grppha} in {\it HEASOFT v 6.32.1} all spectra were grouped with at least 5 counts per energy bin due to low data quality, and the C-statistic was used for fitting to source spectra, but no acceptable models could be achieved for all ULXs. Therefore, unabsorbed flux values were calculated with a generic spectral shape assuming a {\it power-law} photon index of $\Gamma$=1.7 \citep{2020MNRAS.491.1260S} and a Galactic absorption component, N$_{H}$=0.03$\times$10$^{22}$ cm$^{-2}$ using the {\it srcflux} task in {\scshape ciao}. \\

To determine the most accurate positions of the ULXs, a stacked image was created using all available \textit{Swift/XRT} observations due to the short exposure times and poor data statistics of the single \textit{Swift/XRT} observations. The \textit{XIMAGE} was used to source detection for a stacked event file with the signal-to-noise threshold of 3-$\sigma$. Accordingly, three \textit{Swift/XRT} sources were identified at the ULX-1 position in the \textit{Chandra} image (see Figure \ref{F:RGB}). The detail of the X-ray properties of ULX-1 in this study is quite limited because the \textit{Swift/XRT} detector does not have good enough spatial resolution to analyze these three sources. In addition, the X-ray source (ULX-5?) is very close to the core of the galaxy NGC 1566, detected only in the \textit{Chandra} image, therefore \textit{Swift/XRT} analyses could not be performed for this source due to insufficient angular resolution. The \textit{Chandra} unabsorbed X-ray luminosity of this source was calculated as (3 $\pm$ 0.18) $\times$ 10$^{39}$ \textit{erg} s$^{-1}$ in 0.3-10 keV energy band by using \textit{srcflux}, hence a name like ULX-5? was chosen. On the other hand, in this study, due to its multi-wavelength properties (for more detail see Section 3.3), ULX-5? was identified as an AGN candidate. Furthermore, a source labeled ULX-4 in this study was found to be at the ULX luminosity level (>10$^{39}$ \textit{erg} s$^{-1}$) using the \textit{Swift/XRT} observations.\\

Pipeline-processed event files from the XRT detector were processed using the \textit{XSELECT} package, as described in the \textit{Swift/XRT} user manual. Source regions of 30 arcsec were used, with larger accompanying background regions of 50 arcsec, and care was taken to exclude other sources in the source and background regions. Background of galaxy light curves were determined from each event file and checked for background flaring and/or contamination by the core of the galaxy which proved to be a problem in all observations. In other words, the emission of the core of NGC 1566 is fully contaminating all observations of the ULXs. Therefore, in order to avoid both contamination and background flaring, the source regions were reduced until the lowest contamination and flaring were achieved. The result of the best possible source regions was determined to be 18 arcsec, which is the angular resolution of the \textit{Swift/XRT} detector. For the background regions, source-free regions were selected according to the stacked image.\\

\begin{figure*}
\begin{center}
\includegraphics[angle=0,scale=0.3]{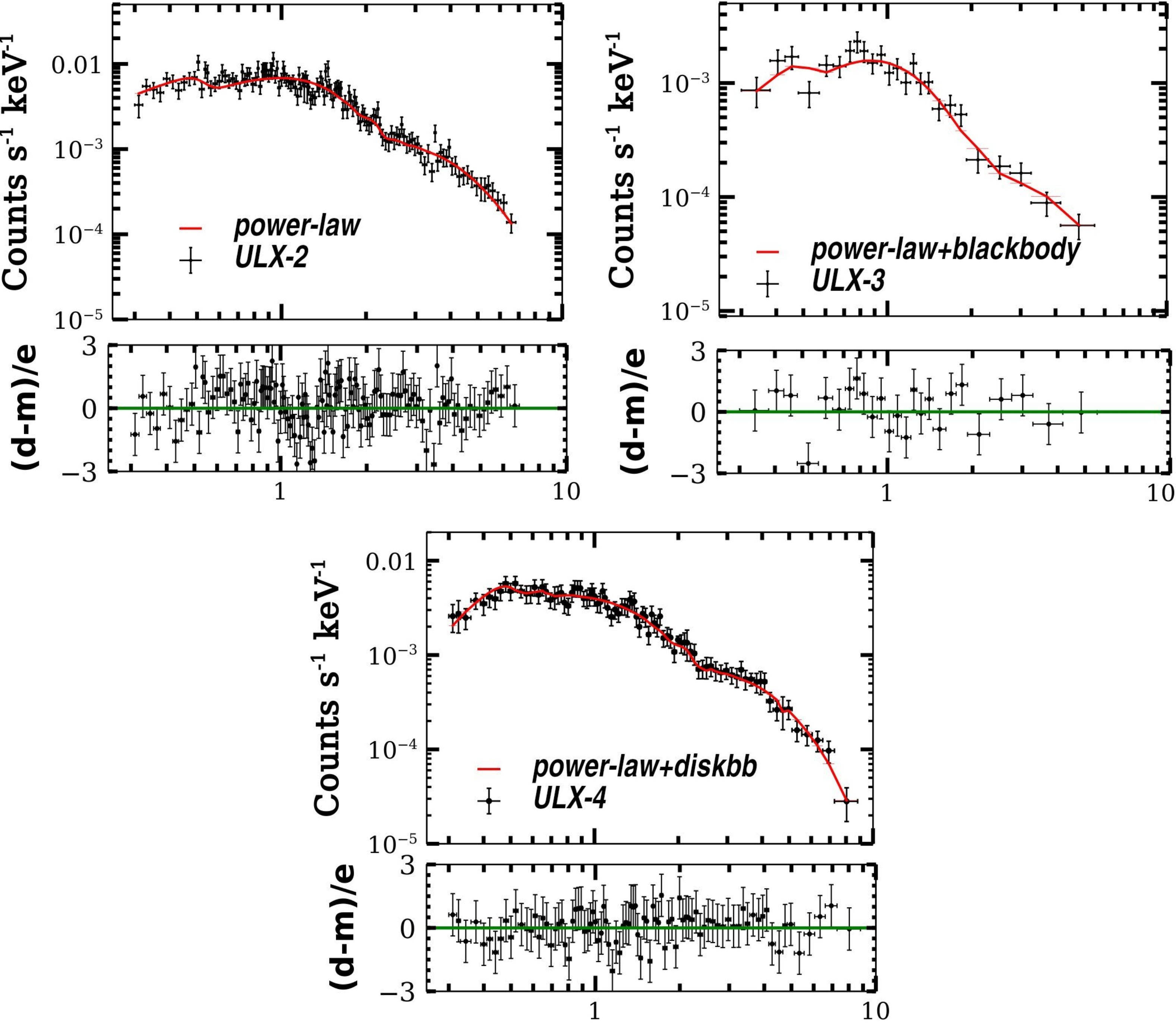}
\caption{Energy spectra for the ULX-2 (left) and ULX-3 (right), and ULX-4 (bottom) using time-averaged {\it Swift/XRT} observations. ((d-m)/e =(data-model)/error).}
\label{F:spectrum}
\end{center}
\end{figure*}

Due to poor data quality, an attempt was made to combine spectra from many observations with similar spectral characteristics, such as hard and soft colors, for fitting purposes, but this attempt was unsuccessful as ULXs could not be detected in each observation. Therefore, time-averaged spectra were obtained for ULX-2, ULX-3, and ULX-4 using all available observations in the 0.3-10 keV energy band. The spectra of the ULXs with the instrument responses and ancillary files were generated for {\it Swift/XRT} observations. According to the averaged source counts, the source energy spectra were grouped using the FTOOLS {\it grppha} with at least 15 counts per energy bin for {\it Swift/XRT} observations. Spectral fits were performed on all available 0.3-10 keV spectral data to determine the best-fitting model for each time-averaged spectrum using {\scshape xspec} v12.13. In all cases a hydrogen column density (\textit{N$_{H}$}) component ({\it tbabs}) was used. ULX-2 was adequately fitted by {\it power-law} (po) and in the case of ULX-3 and ULX-4, the time-averaged spectrum gave better statistics according to the F-test for the {\it power-law+blackbody} and {\it power-law+diskbb} models, respectively at 3-$\sigma$, along with well-constrained model parameters and $\chi^2_{\nu}$ statistics. The best-fit results are presented in Table \ref{T:fit} and Figure \ref{F:spectrum} shows the spectral fits of the ULXs. \\

\begin{table*}
\centering
\begin{minipage}[b]{0.9\linewidth}
\caption{The best model parameters of three ULXs from the time-averaged spectra of {\it Swift/XRT}}
\begin{tabular}{c c c c c c c r r r l }
\hline
Source & N$_{H_{intrinsic}}$ & N$_{\mathrm{{\it po}}}$ & N$_{\mathrm{{\it bb}}}$ & N$_{\mathrm{{\it diskbb}}}$ & $\Gamma$ & {\it Tin/kT} & F$_{X_{unabs}}$ & L$_{X_{unabs}}$ & $\chi^{2}$/dof \\
& (1) & (2) & (3) & (4) & (5) & (6) & (7) & (8) & (9)\\
\hline
ULX-2 & $0.03_{-0.01}^{+0.01}$ & $0.11_{-0.01}^{+0.01}$ & ... & ... & $1.72_{-0.07}^{+0.07}$ & ... & $6.48_{+0.22}^{-0.24}$ & $20.2_{-2.7}^{+2.9}$ & 146.55/132 (1.1)\\
ULX-3 & $0.04_{-0.01}^{+0.01}$ & $1.30_{-0.60}^{+0.61}$ &$3.90_{-0.31}^{+0.81}$ & ... & $2.12_{-0.11}^{+0.12}$ & $0.12_{-0.04}^{+0.06}$ & $1.22_{+0.20}^{-0.22}$ & $4.20_{-0.50}^{+0.60}$ & 20.59/20 (1.0)\\
ULX-4 & $0.05_{-0.01}^{+0.02}$ & $5.78_{-0.38}^{+0.48}$ & ... & $5.36_{-0.48}^{+0.56}$ & $1.55_{-0.11}^{+0.11}$ & $0.26_{-0.06}^{+0.08}$ & $2.78_{+0.26}^{-0.31}$ & $8.64_{-1.8}^{+2.2}$ & 74.63/84 (0.9)\\
\hline
\end{tabular}
\\ Notes. — Col. (1): Intrinsic X-ray absorption value in units of 10$^{22}cm^{-2}$. Col. (2): Normalization parameter of \textit{power-law} model in units of 10$^{-5}$. Col. (3): Normalization parameter of \textit{blackbody} model in units of 10$^{-6}$. Col. (4): Normalization parameter of \textit{diskbb} model in units of 10$^{-6}$. Col. (5): Photon index from the {\it power-law} model. Col. (6): Tin is the temperature of the \textit{Multi-color disk blackbody (diskbb)} in keV (for ULX-4) and kT is the temperature of the {\it blackbody} in keV (for ULX-3). Col. (7): Unabsorbed fluxes in units of 10$^{-13}$ ergs $cm^{-2}$ $s^{-1}$. Col. (8): Unabsorbed luminosities in units of 10$^{39}$ ergs $s^{-1}$ in the 0.3–10 keV energy band, adopting a distance of 17.69 Mpc. Col. (9): The reduced $\chi^{2}$ is given in parentheses. All errors are at the confidence range of 2.706.\\
\label{T:fit}
\end{minipage}
\end{table*}

\subsubsection{Long-term Variability}

\begin{figure*}
\begin{center}
\includegraphics[angle=0,scale=0.4]{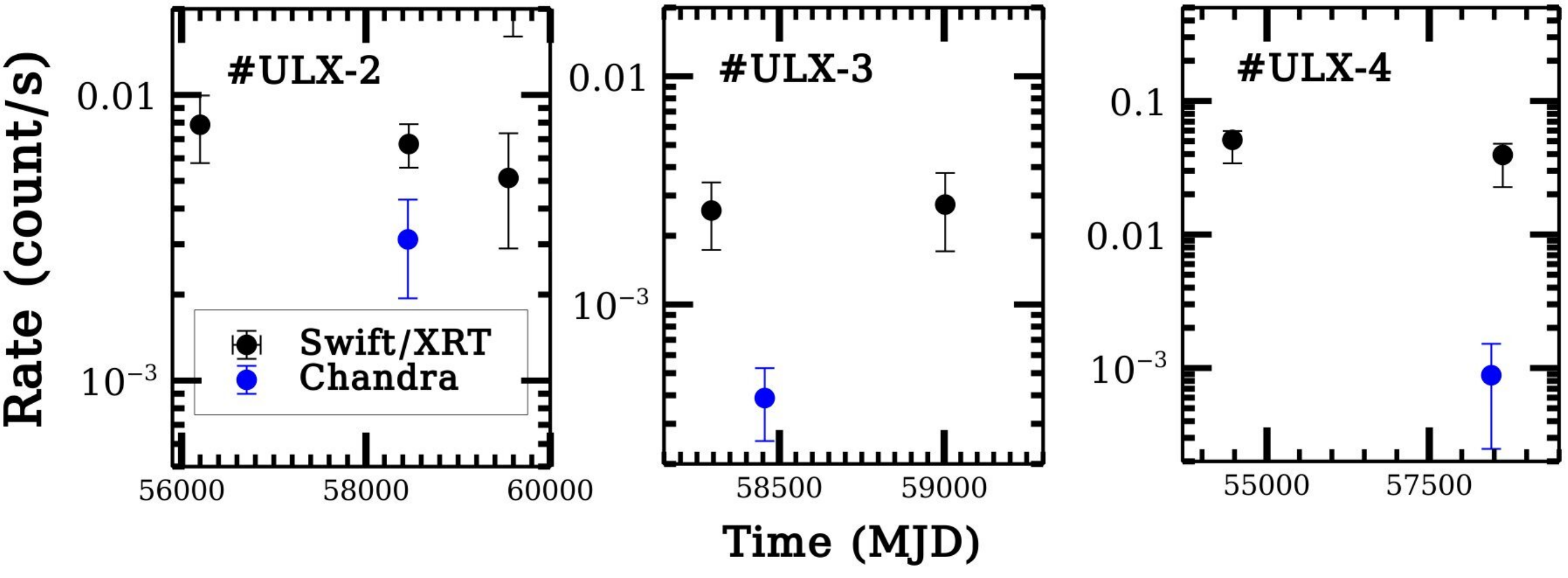}
\caption{Light-curves of the ULX-2 (left), ULX-3 (middle), and ULX-4 (right). The count rates in the energy band 0.3-10 keV were derived by using {\it Swift/XRT} PC mode observations (filled black circles) and \textit{Chandra} data(filled blue circles). }
\label{long}
\end{center}
\end{figure*}

The count rates for {\it Swift/XRT} in PC (Photon-counting) mode were extracted using automated procedures specified on the web page\footnote{https://www.swift.ac.uk/user\_objects/} in the 0.3-10 keV energy band, taking into account the positions determined from the combined event file. Due to the angular resolution of \textit{Swift/XRT} (18 arcsec) and the angular separation between the ULXs and the core of NGC 1566 (> 60 arcsec), the source regions had to be chosen carefully to avoid contamination. Since this automated procedure generally uses regions with large radii when calculating the count rates of the sources, the previously detected variable background flaring was predominantly detected for each source. Therefore, contamination was limited (as much as possible) by choosing a search radius of 18 arcsec using the centroid function, and the binning method was chosen as one bin per observation. However, out of 262 observations, the background subtracted source count rates of ULX-2 were detected in only three observations, while ULX-3 and ULX-4 were detected only in two observations. Due to fact that the majority of single \textit{Swift/XRT} observations have short exposure times, the upper limits (3-$\sigma$) of non-detection were not calculated. There are insufficient high-quality data sets available to place certain constraints on the long-term characteristics of ULXs, but by examining the observations where ULXs are detected, clues for the long-term variabilities may be found. Therefore, according to the \textit{Swift/XRT}, the \textit{Chandra} count rates were normalized with \textit{Chandra PIMMS}\footnote{https://cxc.harvard.edu/toolkit/pimms.jsp} by using a {\it power-law} photon index of $\Gamma$=1.7 and a Galactic absorption component, N$_{H}$=0.03$\times$10$^{22}$ cm$^{-2}$. The long-term light curves of ULX-2, ULX-3, and ULX-4 are plotted in Figure \ref{long}. 

\begin{figure}
\begin{center}
\includegraphics[angle=0,scale=0.3]{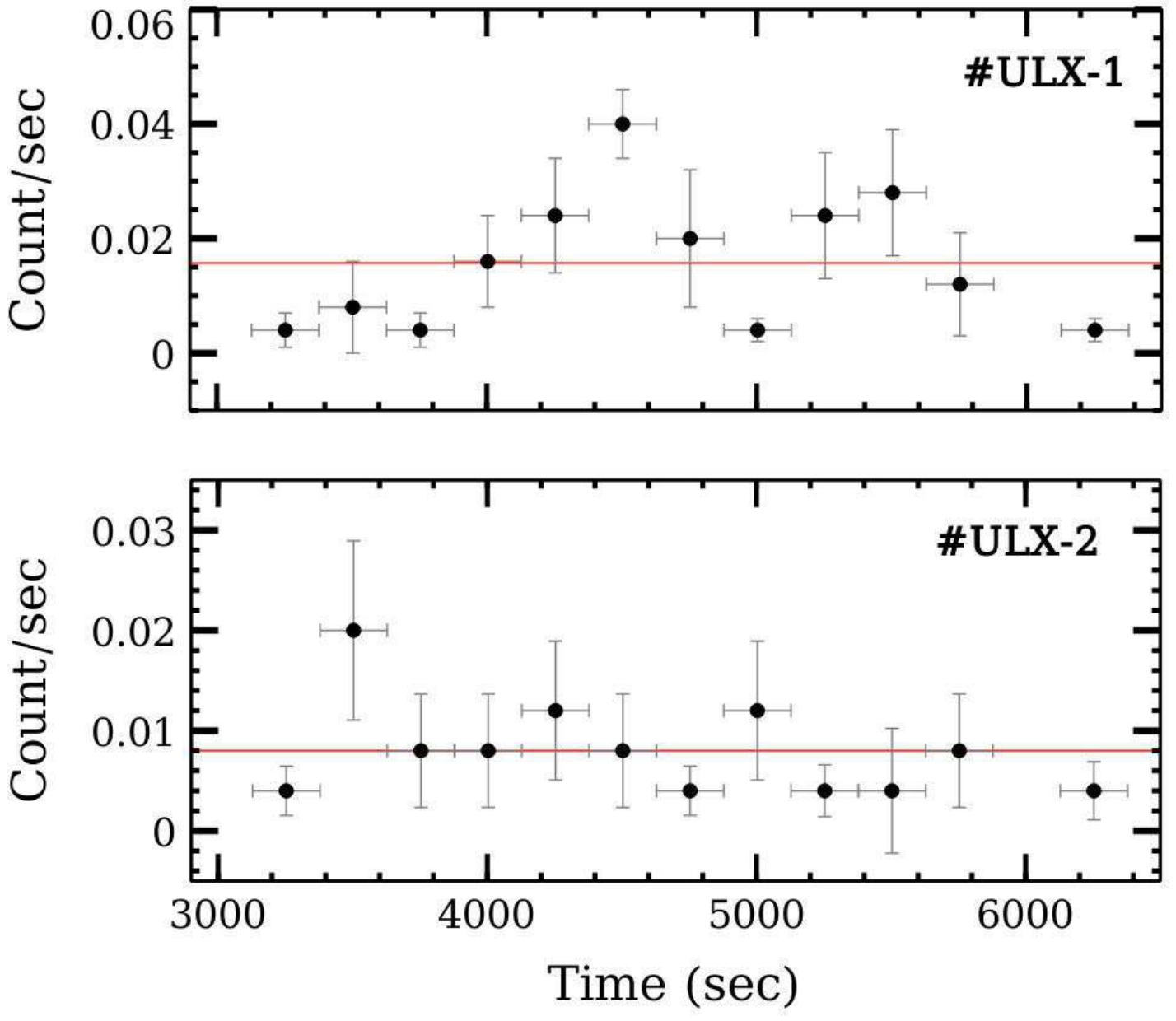}
\caption{Light curves of ULX-1 (top) and ULX-2 (bottom). The bin time in both light curves is set to 250 s in the 0.3-10 keV energy band. In both panels, the red solid line indicates the mean value of the {\it Chandra} X-ray count rate.}
\label{lch}
\end{center}
\end{figure}

To quantify the variability factors (F$_{V}$), the ratios of maximum to minimum count rates (F$_{max}$/F$_{min}$) were derived. Where F$_{max}$ and F$_{min}$ are the maximum and minimum count rates. Considering the variability factors, no significant variability was observed in the \textit{Swift/XRT} observations, whereas when the \textit{Chandra} and \textit{Swift/XRT} observations were compared, the variability factors were derived as $\sim$ 7 $\pm$ 2 and 60 $\pm$ 11 for ULX-3 and ULX-4, respectively. In the case of ULX-2, no significant variation could be found (2 $\pm$ 1). Finally, the short-term count rate variability was also searched for all ULXs using {\it Chandra} data (3 ks). For this, the light curve of the ULXs was binned over intervals of 100 s and 250 s in the 0.3$-$10 keV energy band using the {\scshape ciao} tool {\it dmextract}. A \textit{barycenter} correction was applied to the data before the timing analysis using the {\scshape ciao} tool {\it axbary}. The short light curves of ULX-1 and ULX-2 are shown in Figure \ref{lch}. F$_{V}$ for the short-time light curves were derived as $\sim$ 10 $\pm$ 4 and 5 $\pm$ 3, respectively. The light curves for ULX-3 and ULX-4 are not plotted due to the low data statistics.\\

\subsection{Optical and Near-infrared data}

\subsubsection{Source Detection \& Photometry}

To identify the positions and aperture photometry of the sources, the steps described in \cite{2023arXiv230611163A} (and references therein) were followed. Sources were detected using $\geq$ 3-$\sigma$ threshold detection. Vega magnitudes were obtained from aperture photometry using a circular aperture with a radius of 3 pixels, and the background was subtracted from an annulus nine pixels from the source center. For all drizzled {\it HST/WFC3} images, point-like sources were detected with the {\it daofind} task, and aperture photometry for these sources was performed using the \textit{DAOPHOT} package \citep{1987PASP...99..191S} in the {\scshape iraf}\footnote{https://iraf-community.github.io/} (\textit{Image Reduction and Analysis Facility}). In the case of \textit{JWST} observations, 3 pixels of aperture radius were selected for the aperture photometry and nine pixels for the background (for more details see \citealp{2023arXiv230611163A}) using {\scshape photutils v1.8}\footnote{https://photutils.readthedocs.io/en/stable/index.html\#}.

\subsection{Determination of Counterparts}

To determine both the IR and optical counterparts of the ULXs precise astrometry was performed by following the works of \cite{2022MNRAS.515.3632A} and \cite{2023arXiv230611163A}. Since no matching sources were identified from the {\it GAIA/DR3} source catalog with the XRB or XRB candidates, the reference sources were searched by comparing {\it Chandra} X-ray image with {\it NIRCam} observations.

\begin{figure}
\begin{center}
\includegraphics[width=\columnwidth]{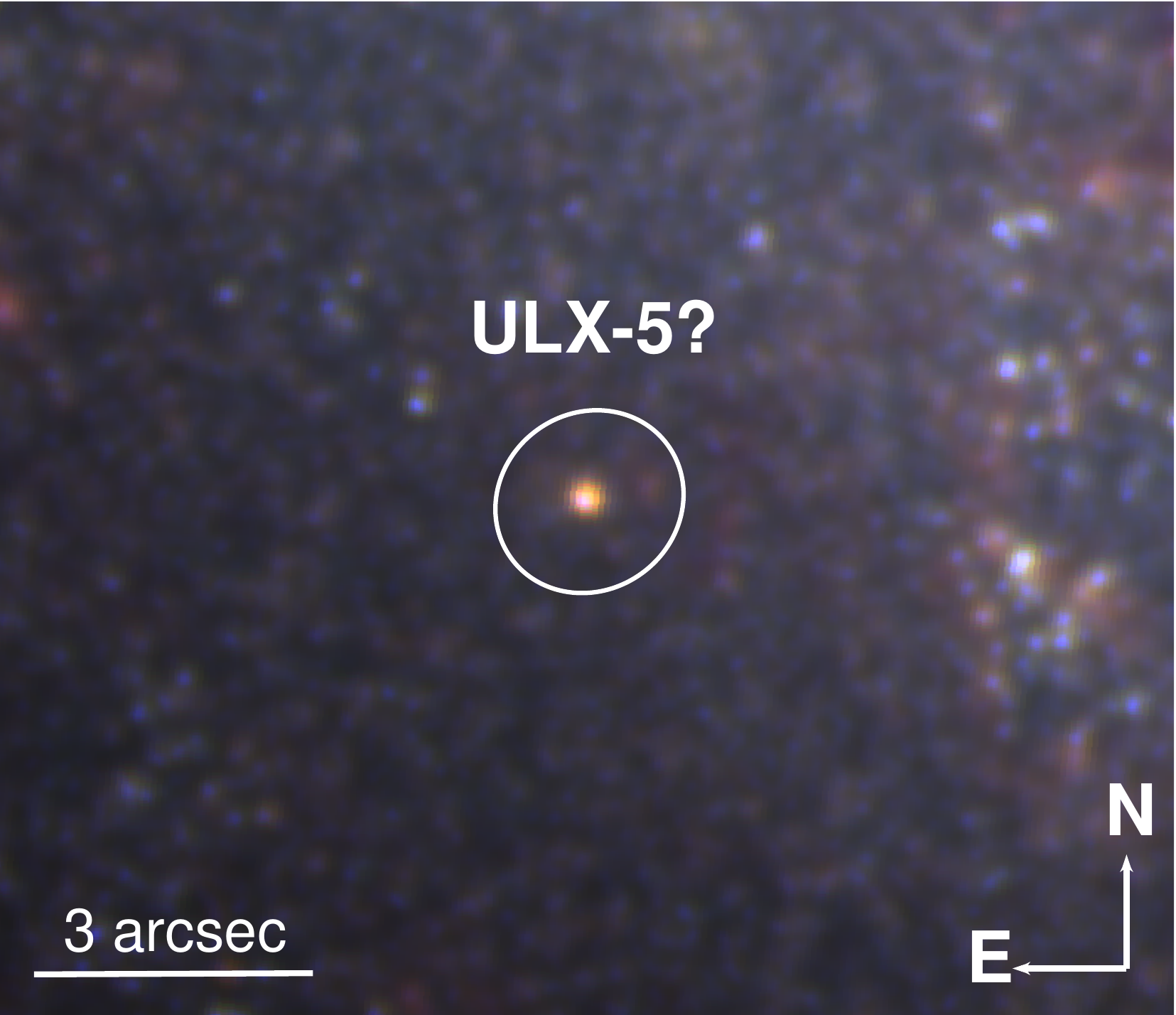}
\caption{\textit{Chandra} X-ray position of ULX-5? on the RGB {\it NIRCam} image. The white circle shows the \textit{Chandra} error ellipse.}
\label{F:ULX5}
\end{center}
\end{figure}

Only four of the eight X-ray sources (including ULXs) are in the field of view of the {\it JWST/NIRCam} images. These XRB candidates were compared with NIR point sources, and two reference sources (excluding ULXs) were found for astrometric calculations. The coordinates in degrees of the identified reference sources are R.A = 65.00385 and Decl. = -54.94534 (AGN) and R.A = 65.00172 Decl. = -54.93789 ( core of the galaxy). The astrometric offsets between the {\it Chandra} and {\it JWST/NIRCam} F200W images were found as -0$\arcsec$.22 $\pm$ 0$\arcsec$.01 for R.A and -0$\arcsec$.04 $\pm$ 0$\arcsec$.05 for Decl. with 1-$\sigma$ errors. The total astrometric errors between {\it Chandra}-{\it JWST/NIRCam} were derived as 0$\arcsec$.12. The astrometric error radius was derived as 0$\arcsec$.38 with a 90\% confidence level. According to the corrected X-ray positions, a unique NIR counterpart for ULX-1, two NIR counterparts for ULX-2 (hereafter 2-a and 2-b), and several NIR counterparts for ULX-4 were identified within the astrometric error radius. A unique source was identified that is too extended to be the donor star within the derived astrometric error radius of ULX-5? (see figure \ref{F:ULX5}). Although it was identified as a potential ULX source in Section 3.1 due to its X-ray luminosity, the multi-wavelength observations suggest that it is more likely to be an uncatalogued AGN. Therefore, the possibility of this source being a ULX is excluded in this study. In the case of ULX-3, the NIR counterpart(s) could not be identified for ULX-3 because it was not observed by {\it JWST}.\\

\begin{figure*}
\begin{center}
\includegraphics[angle=0,scale=0.25]{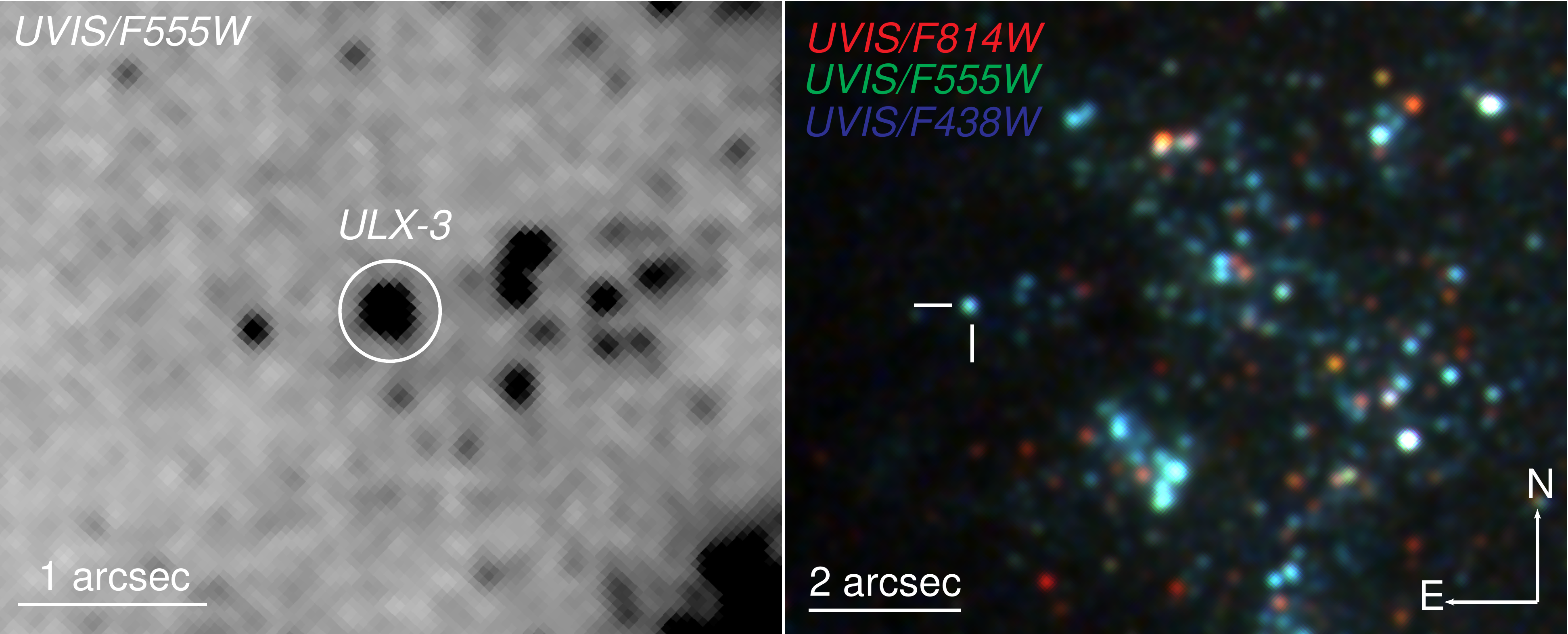}
\caption{{\it HST/WFC3} F555W (left) and RGB (right) images of the corrected X-ray position of ULX-3. A white circle and white bars indicate the position of ULX-3. The images are smoothed with a 3 arcsec Gaussian. In two panels the north is up.}
\label{F:ulx3}
\end{center}
\end{figure*}

\begin{figure*}
\begin{center}
\includegraphics[angle=0,scale=0.27]{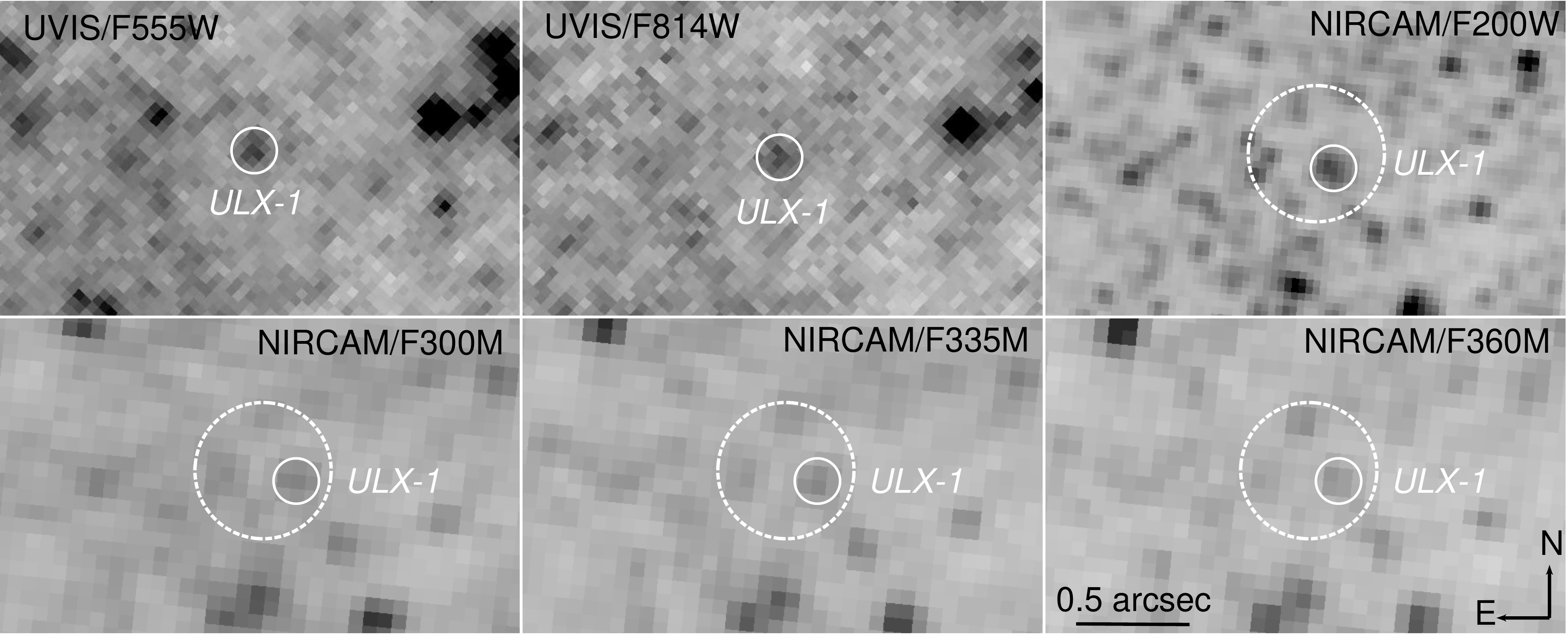}
\caption{The corrected X-ray position of ULX-1 is shown on different images from {\it HST/WFC3} and {\it JWST/NIRCam}. The dashed and solid white circles represent the astrometric error radius and the positions of the counterparts. All panels have the same scale.}
\label{F:ulx1}
\end{center}
\end{figure*}

\begin{figure*}
\begin{center}
\includegraphics[angle=0,scale=0.27]{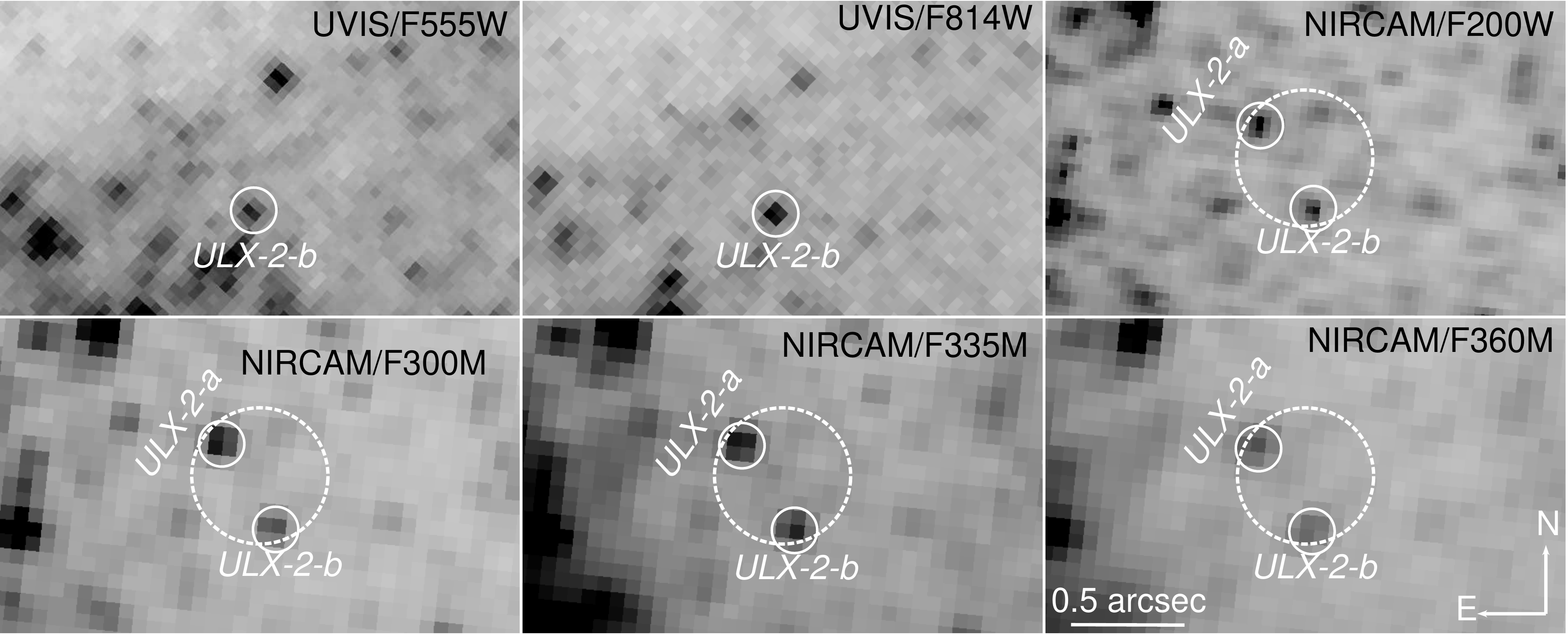}
\caption{The corrected X-ray position of ULX-2 is shown on different images of {\it HST/WFC3} and {\it JWST/NIRCam}. Other explanations are given as in Figure \ref{F:ulx1}.}
\label{F:ulx2}
\end{center}
\end{figure*}

\begin{figure*}
\begin{center}
\includegraphics[angle=0,scale=0.25]{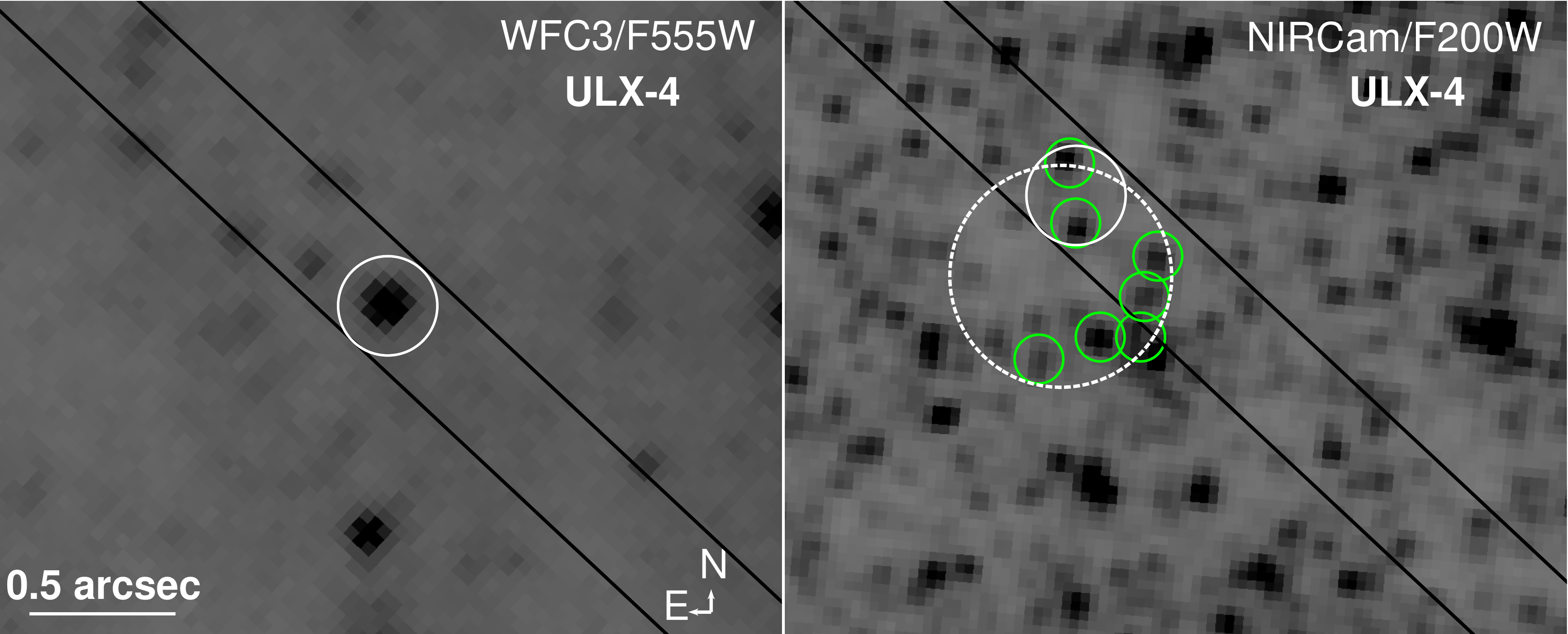}
\caption{ Corrected X-ray positions of ULX-4 on the {\it HST/WFC3} F555W (left) and \textit{JWST/NIRCam} F200W (right) images. The dashed white circle shows the astrometric error radius of 0.38 arcsec and green circles show NIR sources detected above the 3-$\sigma$ threshold. The solid white circles represent the optical source.}
\label{F:ulx4}
\end{center}
\end{figure*}

Moreover, relative astrometry was performed between {\it JWST/NIRCam} and {\it HST/WFC3} images to determine the optical counterparts of ULXs using the {\it GAIA/DR3} source catalog. The astrometric offsets between \textit{JWST/NIRCam} and {\it HST/WFC3} were derived as R.A= 0$\arcsec$.26 $\pm$ 0$\arcsec$.01 and Decl.=0$\arcsec$.33 $\pm$ 0$\arcsec$.01 at 1-$\sigma$ significance. At the corrected position of the NIR counterpart, a unique optical source was identified for ULX-1, while only a unique optical source consistent with 2-b was identified for ULX-2 at the 3-$\sigma$ detection threshold only in the \textit{HST/WFC3} F555W and F814W images. In addition, a unique optical counterpart was identified for the ULX-3. Figure \ref{F:ulx3} shows the corrected X-ray position of the ULX-3 on the F555W and RGB {\it HST/WFC3} images. The corrected X-ray positions of the counterparts of ULX-1, ULX-2, and ULX-4 are shown in Figures \ref{F:ulx1}, \ref{F:ulx2}, and \ref{F:ulx4}. The \textit{Chandra} and corrected X-ray positions of counterparts are given in Table \ref{T3}. The dereddened Vega magnitudes, corrected with A$_{V}$ $=$ 0.04 mag \citep{2022ApJ...925..101T}, of the optical and NIR counterparts are given in Table \ref{T:fotometry}. 

\begin{table*}
\centering
\caption{{\it Chandra} and corrected X-ray positions ULXs for {\it JWST/NIRCam} and {\it HST} images.}
\begin{tabular}{cccrccccccc}
\hline
Source number & {\it Chandra} R.A.& {\it Chandra} Decl.& {\it JWST} R.A.& {\it JWST} Decl. & {\it HST} R.A.& {\it HST} Decl. \\
... & (hh:mm:ss.sss) & ($\degr$ : $\arcmin$ : $\arcsec$) & (hh:mm:ss.sss) & ($\degr$ : $\arcmin$ : $\arcsec$) & (hh:mm:ss.sss) & ($\degr$ : $\arcmin$ : $\arcsec$)\\
\hline
ULX-1 & 4:19:56.08 & -54:56:36.5& 4:19:56.11 & -54:56:36.4 & 4:19:56.07 & -54:56:36.8\\
ULX-2 & 4:19:56.55 & -54:55:29.3& 4:19:56.58 & -54:55:29.2 & 4:19:56.54 & -54:55:29.6 \\
ULX-3 & 4:20:10.12 & -54:56:42.0& ... & ... & 4:20:10.06 & -54:56:42.31 \\
ULX-4 & 4:20:05.03 & -54:56:56.8& 4:19:56.58 & -54:55:29.2 & 4:20:05.01 & -54:56:57.1 \\
ULX-5? &4:20:05.00 & -54:56:57.4 &... & ... & ... & ... \\
\hline
\label{T3}
\end{tabular}
\\Note: The coordinates of the NIR and optical counterparts are given according to their positions in the F200W and F555W images.\\
\end{table*}

\begin{table*}
\centering
\caption{The derived dereddened Vega magnitudes of the optical and NIR counterparts.}
\begin{tabular}{cccccccccc}
\hline
Counterparts & UVIS/F555W & UVIS/F814W & NIRCAM/F200W & NIRCAM/F300M & NIRCAM/F335M & NIRCAM/F360M \\
\hline
{\it ULX-1} & 25.87 $\pm$ 0.12 & 24.81 $\pm$ 0.11 & 21.92 $\pm$ 0.08 & 22.65 $\pm$ 0.08 & 22.51 $\pm$ 0.06 & 22.53 $\pm$ 0.05 \\
{\it ULX-2-a} & >28 & >26.8 & 22.84 $\pm$ 0.06 & 20.74 $\pm$ 0.06 & 20.73 $\pm$ 0.05 & 20.68 $\pm$ 0.04 &\\
{\it ULX-2-b} & 25.36 $\pm$ 0.16 & 23.94 $\pm$ 0.08 & 22.12 $\pm$ 0.06 & 21.65 $\pm$ 0.06 & 21.42 $\pm$ 0.05 & 20.77 $\pm$ 0.06 \\
\\
& UVIS/F275W & UVIS/F336W & UVIS/F438W & UVIS/F555W & UVIS/F814W \\
{\it ULX-3} & 22.39 $\pm$ 0.03 & 22.68 $\pm$ 0.03 & 24.16 $\pm$ 0.03 & 24.12 $\pm$ 0.02 & 24.03 $\pm$ 0.03 \\
\hline
\end{tabular}
\label{T:fotometry}
\end{table*}

\subsection{Spectral Energy Distributions}

\begin{figure}
\begin{center}
\includegraphics[width=\columnwidth]{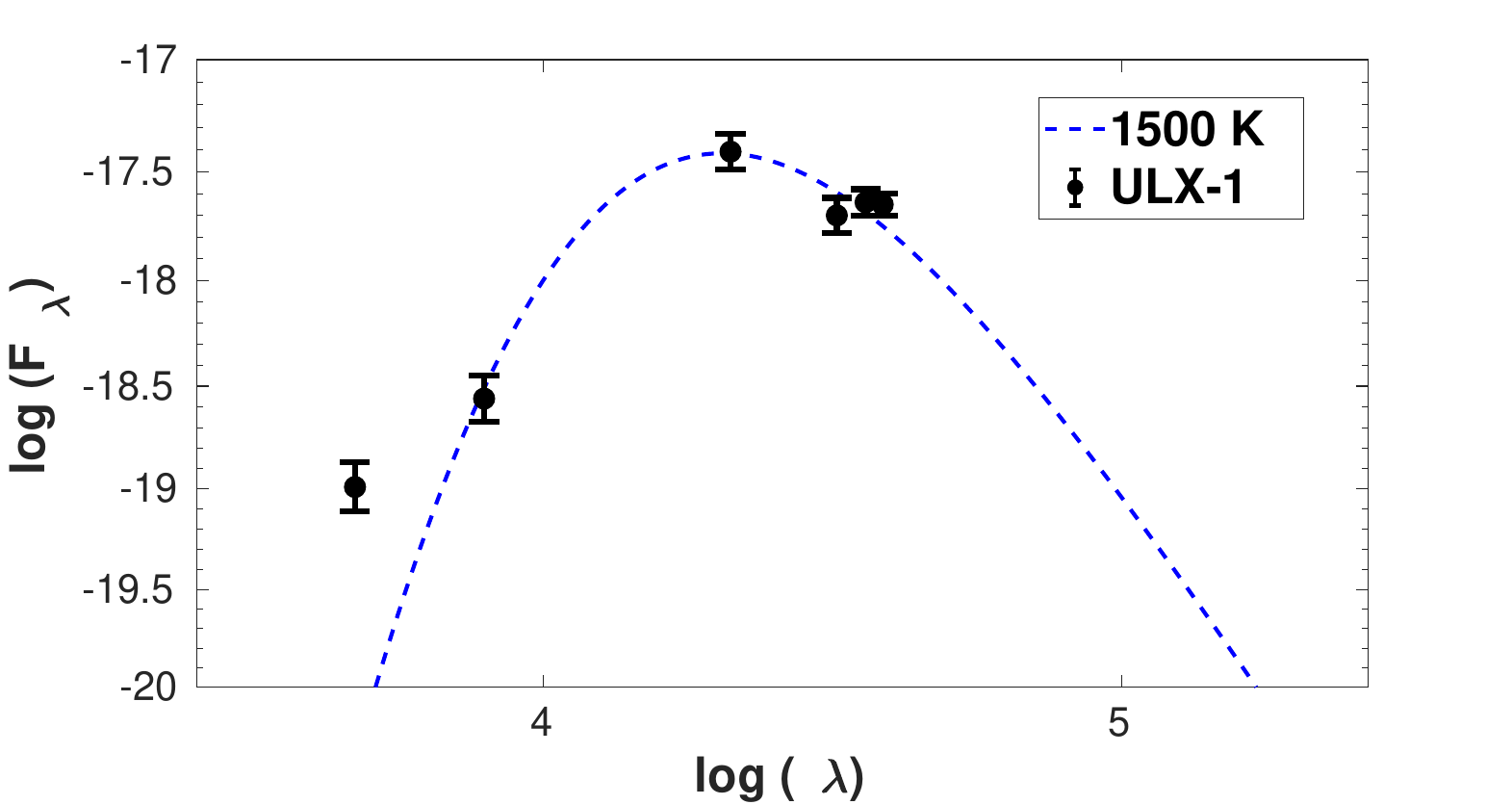}
\includegraphics[width=\columnwidth]{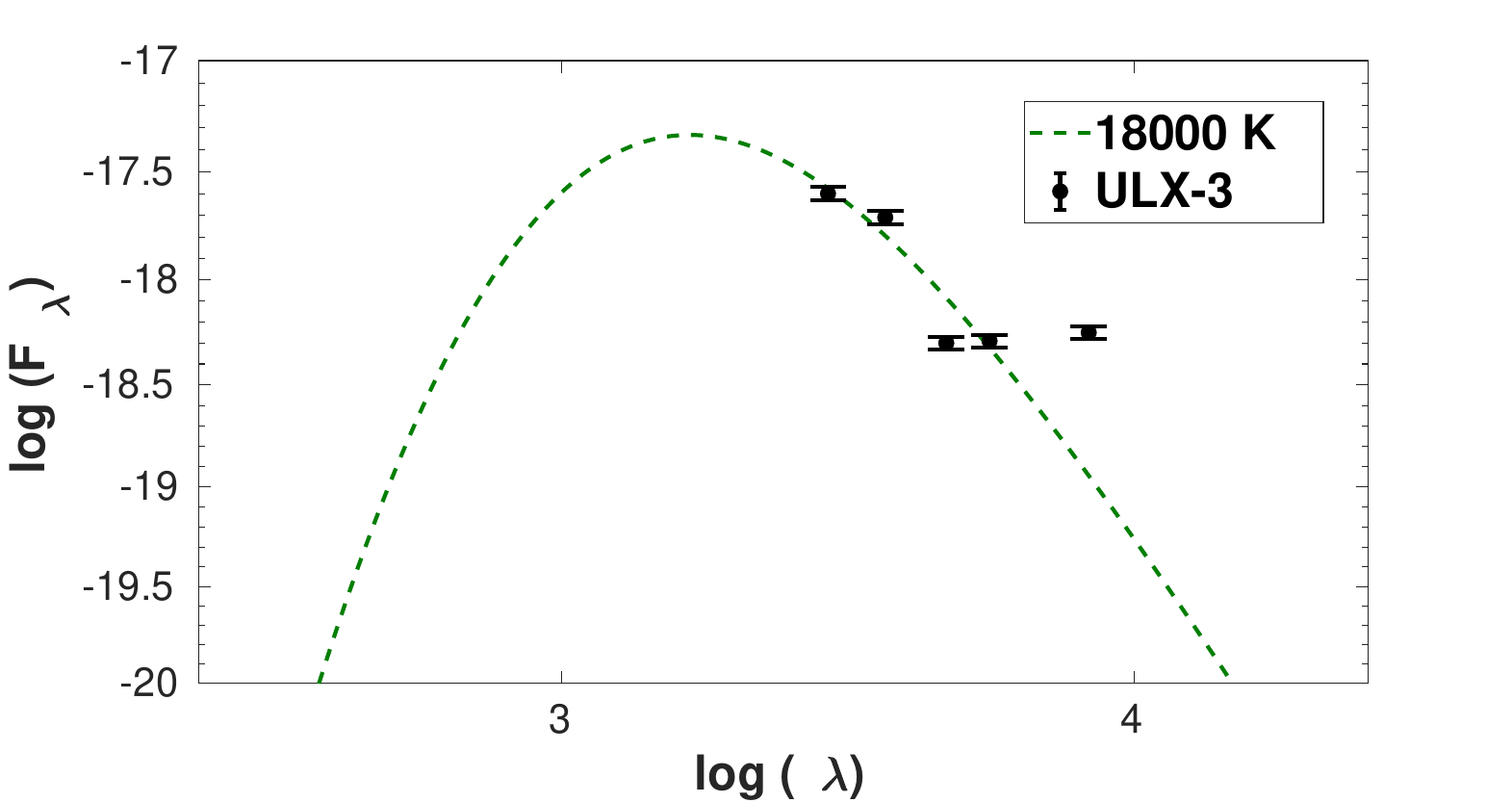}
\caption{The dereddening corrected SEDs for ULX-1 (top) and for ULX-3 (bottom). The {\it blackbody} models for ULX-1 and for ULX-3 are shown by blue and green dashed lines. All data are shown with filled black circles with their respective errors with bars. The units of y and x axes are erg s$^{-1}$ cm$^{-2}$ \AA$^{-1}$ and \AA, respectively.}
\label{F:SED1}
\end{center}
\end{figure}

The SEDs (spectral energy distributions) of the counterparts were plotted by using the flux values derived from Table \ref{T:fotometry} to constrain the optical and NIR emission of counterparts. For this, the SEDs were fitted using either a {\it blackbody} or a {\it power-law}, F$\propto$ $\lambda^{\alpha}$, spectrum by following the similar approach given in our previous study of \cite{2022MNRAS.515.3632A}. The reddening corrected SEDs of ULX-1 and ULX-3 were adequately well-fitted by the {\it blackbody} with temperature ({\it T}) of $\sim$ 1500 K and $\sim$ 18000 K, respectively (see Figure \ref{F:SED1}). In the case of ULX-2, reddening corrected SED of 2-a was adequately well-fitted by a {\it blackbody} model with \textit{T}=700 $\pm$ 120 K, and SED of ULX-2-b was well-fitted by a {\it power-law} model with $\alpha$=1.19 $\pm$ 0.10 at 3-$\sigma$ confidence level. The reduced chi-square, $\chi^2_{\nu}$, values of ULX-1, 2-a, 2-b, and ULX-3 are 0.70, 0.55, 0.67 and 0.81, respectively. For ULX-2, the number of degrees of freedom (dof) is four while in the case of both ULX-1 and ULX-3 is three.

\begin{figure}
\begin{center}
\includegraphics[width=\columnwidth]{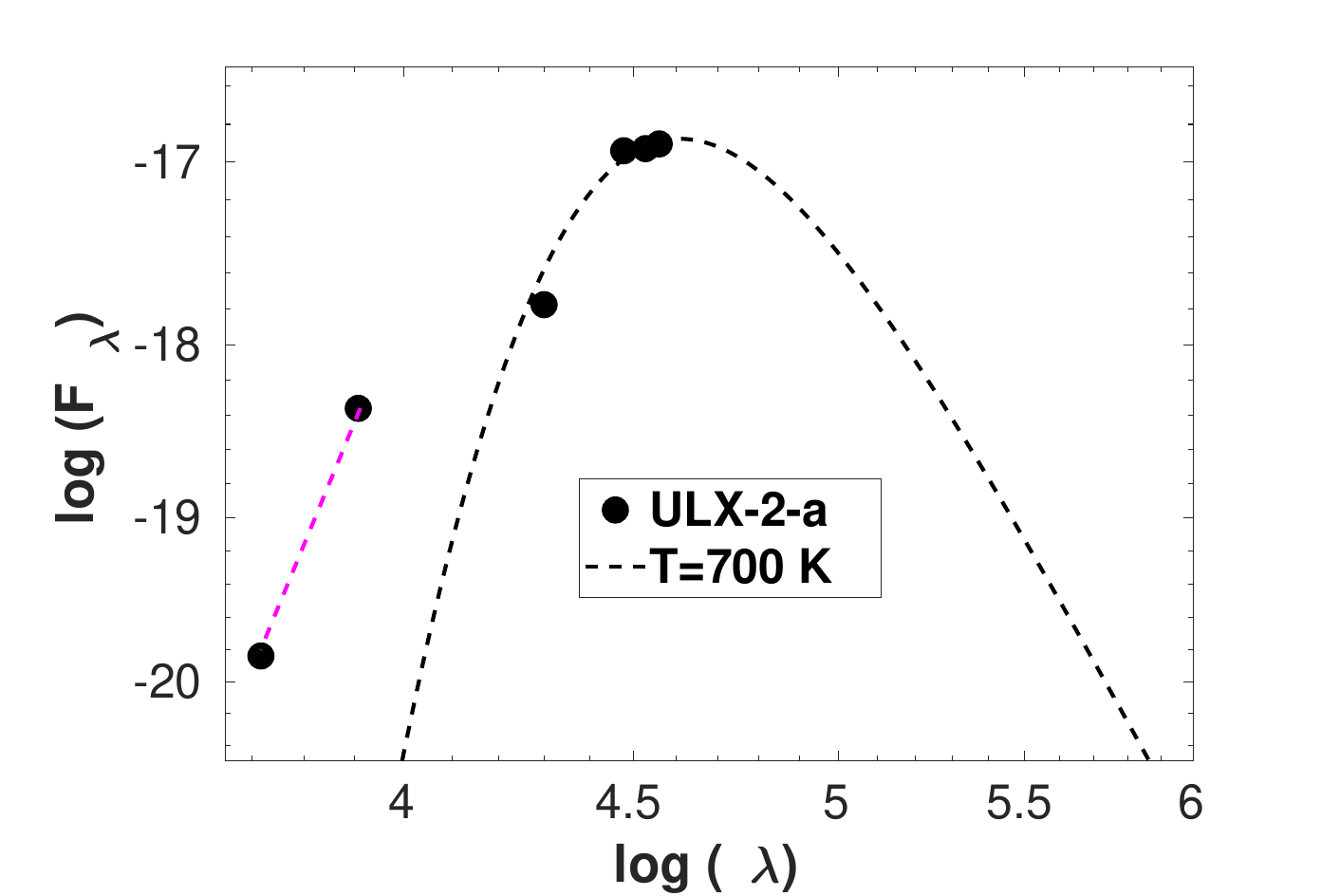}
\includegraphics[width=\columnwidth]{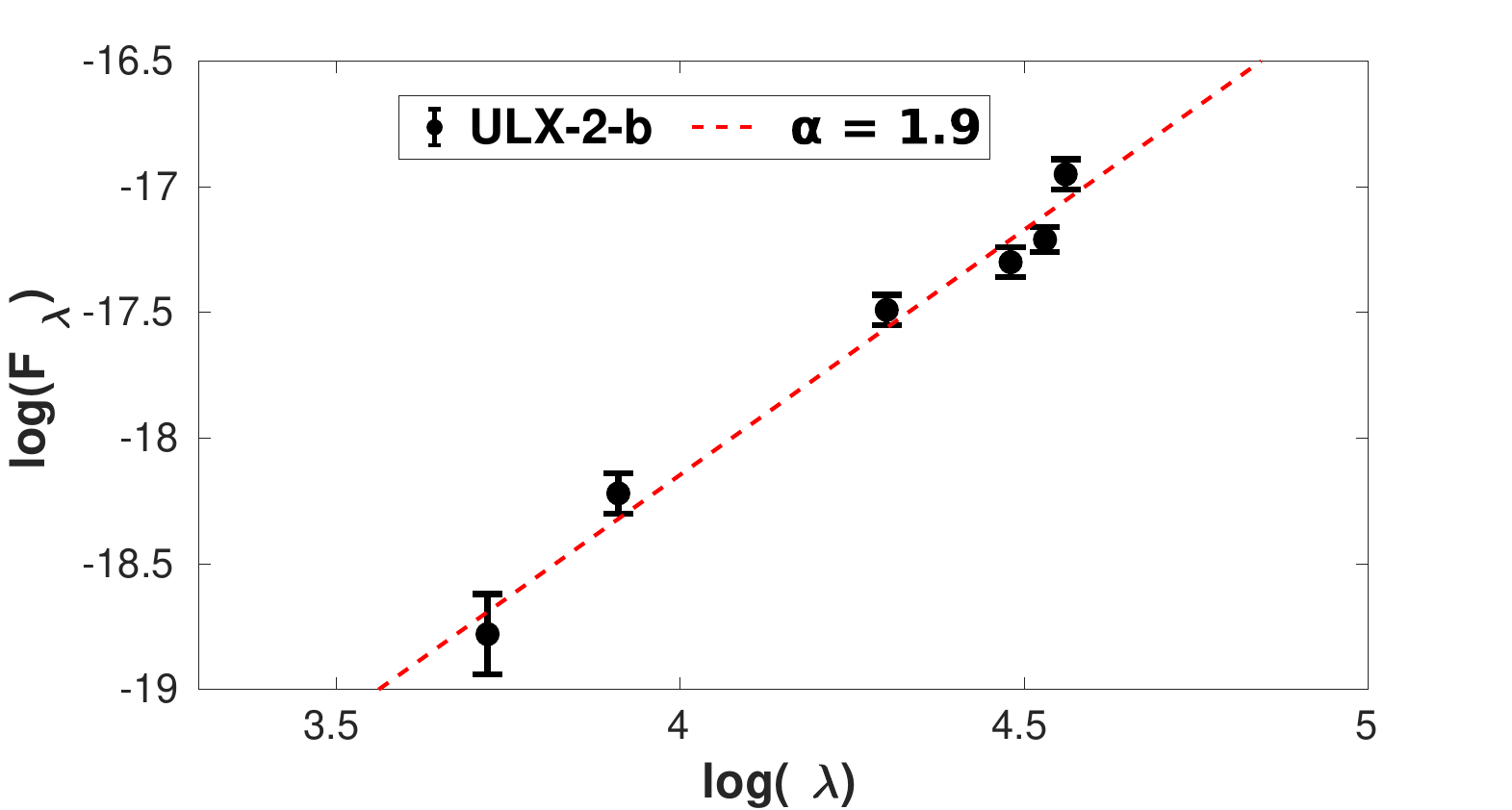}
\caption{The dereddening corrected SEDs of 2-a (top) and 2-b (bottom). The {\it power-law} and \textit{blackbody} models are shown by the gray and red dashed lines. Other explanations are the same as given in Figure \ref{F:SED1}}
\label{F:SED2}
\end{center}
\end{figure}

\section{Results and Discussion} \label{sec:4}

\subsection{X-ray}

\textit{ULX-1}: As can be seen in Figure \ref{F:RGB}, two X-ray sources are located close to ULX-1 in the \textit{Swift/XRT} image. However, these sources were not detected in the \textit{Chandra} image, so they could be candidates for transient X-ray sources. Therefore, a detailed study of the \textit{Swift/XRT} X-ray properties of ULX-1 could not be performed due to insufficient angular resolution. On the other hand, the \textit{Chandra} X-ray luminosity of ULX-1 was derived as $\sim$ (6 $\pm$ 0.12) $\times$ 10$^{39}$ \textit{erg} s$^{-1}$ using \textit{srcflux} in the 0.3-10 keV energy band. It also varies by almost an order of magnitude in the short-term (3ks) light curve constructed from the \textit{Chandra} observation. Due to the lack of high spatial resolution images such as \textit{Chandra}, the long-term variability feature is not clear. Based on a distance of 17.69 Mpc, \textit{ROSAT/HRI} \citep{2005ApJS..157...59L} luminosity values are re-derived as (3-10) $\times$ 10$^{39}$ \textit{erg} s$^{-1}$. On the other hand, the spatial resolution of \textit{ROSAT/HRI} (2 arcsec/pixel) is almost the same as \textit{Swift/XRT} (2.36 arcsec/pixel), probably contributed by the potential transient X-ray source candidates shown in Figure \ref{F:RGB}.\\

\textit{ULX-2}: Time-averaged 0.3-10 keV \textit{Swift/XRT} spectrum of the ULX-2 is composed of \textit{power-law} with \textit{$\Gamma$} = 1.72 \textit{a} non-thermal emission continuum due to Comptonization in corona. From the best fitting \textit{tbabs} $\times$ \textit{power-law} model, the unabsorbed X-ray luminosity (\textit{L$_{X}$}) of ULX-2 was derived as $\sim$ 10$^{40}$ \textit{erg} s$^{-1}$ which is compatible with the luminosity values of \textit{ROSAT/HRI} (3-10) $\times$ 10$^{39}$ \textit{erg} s$^{-1}$ using distance of 17.69 Mpc. Outbursts for the typically Galactic X-ray binaries, begin and end in the hard state which shows a power-law X-ray spectrum with a hard photon index (1.4 < \textit{$\Gamma$} < 2.1) and also strong variability on short timescales. However, in the case of ULX-2, there was no clear evidence of an outburst due to low data quality. In addition, no clear evidence for short-term count rate variability was found. For the properties of short and long-term variability, high-quality X-ray observations with good enough resolution, such as the {\it Chandra} observatory, are needed. \\

\begin{figure}
\begin{center}
\includegraphics[angle=0,scale=0.35]{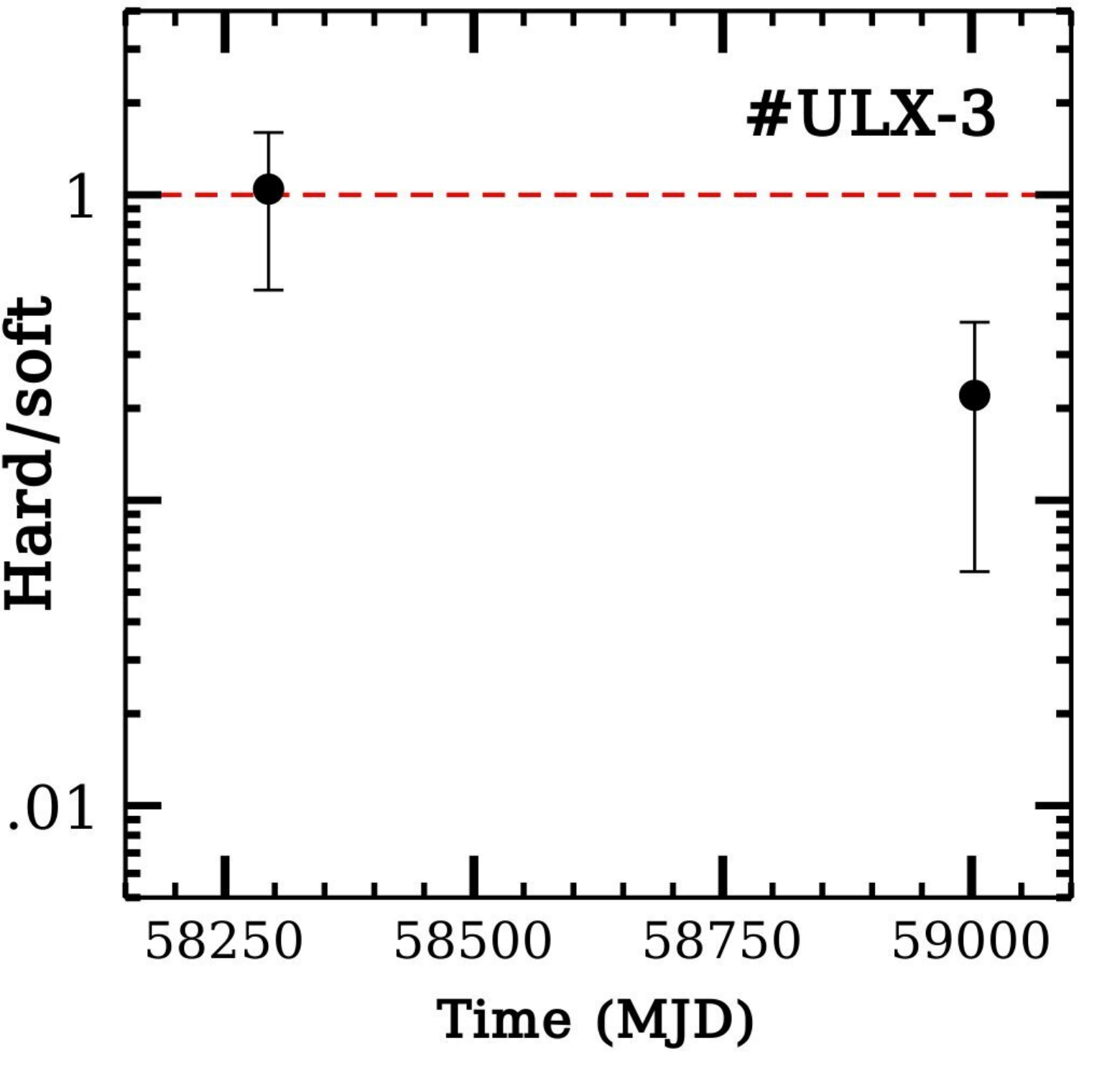}
\caption{The \textit{Swift/XRT} spectral hardness ratios (hard/soft) of ULX-3. The hardness ratios were derived in two energy bands, soft = 0.3-1.5 keV and hard = 2-8 keV.}
\label{F:hardness}
\end{center}
\end{figure}

\textit{ULX-3}: The time-averaged \textit{Swift/XRT} spectrum of the ULX-3 is well-fitted by two-component \textit{power-law + blackbody} model with \textit{$\Gamma$} $\sim$ 2 and \textit{kT} = 0.12 keV. The unabsorbed X-ray luminosity was derived as 4 $\times$ 10$^{39}$ \textit{erg} s$^{-1}$ in 0.3-10 keV energy band. The {\it blackbody} component is dominant and contributes to 80\% of the luminosity. The {\it power-law} component represents a hard tail in the spectrum (out of a standard accretion disk) while a single cool \textit{blackbody} component may indicate the soft thermal emission arises from the photosphere of thick outflows (\citealp{2016ApJ...831..117F} and references therein). To search for a possible spectral state transition, hardness ratios, defined as Hard/Soft (Hard=2-8 keV, Soft= 0.3-1.5 keV) were plotted using two \textit{Swift/XRT} observations. As can be seen from the diagram of the hardness ratios in Figure \ref{F:hardness}, a clue of transition from hard to soft states is observed for ULX-3, as seen in typical high-mass Galactic X-ray binaries \citep{2017ARA&A..55..303K}. A scenario for this state transition is that the accretion rate increases, more mass falls into the accretion disk, which can lead to more soft X-ray emission. In other words, the mass transfer and the change in disk geometry can be observed as transitions. Provided that \textit{Chandra} and \textit{Swift/XRT} observations are compared, as seen in Figure \ref{long}, the possibility of long-term variability is not ruled out, but the short-term variability of ULX-3 is not fully clear due to a lack of high-quality data.\\

\textit{ULX-4}: The time-averaged spectrum of the ULX-4 is well-fitted by a two-component model \textit{power-law + diskbb} with \textit{$\Gamma$} $\sim$ 1.5 and \textit{Tin} = 0.26 keV, and unabsorbed X-ray luminosity of $\sim$ 10$^{40}$ \textit{erg} s$^{-1}$ in 0.3-10 keV energy band. ULX-4 has not been identified as ULX in previous studies and/or observations but, this luminosity value makes it a strong ULX candidate. The X-ray spectrum of accreting black hole binaries (BHBs) is commonly modeled using a combination of a \textit{diskbb} and a power-law (e.g., \citealp{2016ApJ...831..117F}). As can be seen from the time-averaged spectrum in Figure \ref{F:spectrum}, the \textit{diskbb} component is dominant below 2 keV, and the \textit{power-law} component is dominant above 2 keV. The \textit{diskbb} component is predominant and carries 65\% of the luminosity. The thermal state of BHB is represented by a spectrum consisting of a dominant thermal \textit{diskbb} with secondary \textit{power-law} or Comptonization \citep{2006ARA&A..44...49R}. However, results of the discovery of NSs in ULX systems (e.g., NGC 7793 P13 \cite{2017MNRAS.466L..48I}) and the availability of high-quality X-ray energy spectra of ULXs (e.g., NGC 5907 ULX-1; \cite{2013MNRAS.434.1702S}) from new generation X-ray observatories, there is no longer a strong suspicion of the presence of IMBHs for ULXs. For this source, the variability factor was found as $\sim$ 60 (see Figure \ref{long}) using both \textit{Chandra} and \textit{Swift/XRT} count rates that this may make it a transient ULX candidate. The short-term variability for ULX-4 is unclear due to lack of quality and/or insufficient observations.\\

\subsection{Astrometry and Identification of Counterparts}

Taking into account astrometric calculations, for ULX-1, ULX-2, and ULX-4 at least one NIR counterpart was identified within the astrometric error radius of 0.38 arcsec, and at least one of the NIR counterparts matches the optical counterpart. As seen in the F200W image of Figure \ref{F:ulx1}, there are faint sources below the 3-$\sigma$ detection threshold within the error radius, except for the bright NIR counterpart. In this study, the source that is bright and close to the center of the error radius is considered as the NIR counterpart, but the possibility that other fainter sources could be possible donor star candidates for ULX-1 is not completely excluded. In the case of ULX-4, seven NIR counterparts were detected within the error radius therefore, there are too many possibilities to decide which is a possible donor star for ULX-4. In addition, these sources could not be resolved in the \textit{JWST/MIRI} observations due to their relatively low angular resolution. Hence, no constraints could be placed on the possible donor star of the new ULX candidate identified in this study. Simultaneous X-ray and multi-wavelength observations are needed to identify possible donor stars. In the case of ULX-3, since it was not observed by \textit{JWST} its NIR counterparts were not determined, and only a unique optical counterpart was identified. In this study, only the optical and IR counterparts of ULX-1, ULX-2, and ULX-3 were analyzed in detail to constrain the nature of possible donor stars. \\

\begin{figure}
\begin{center}
\includegraphics[width=\columnwidth]{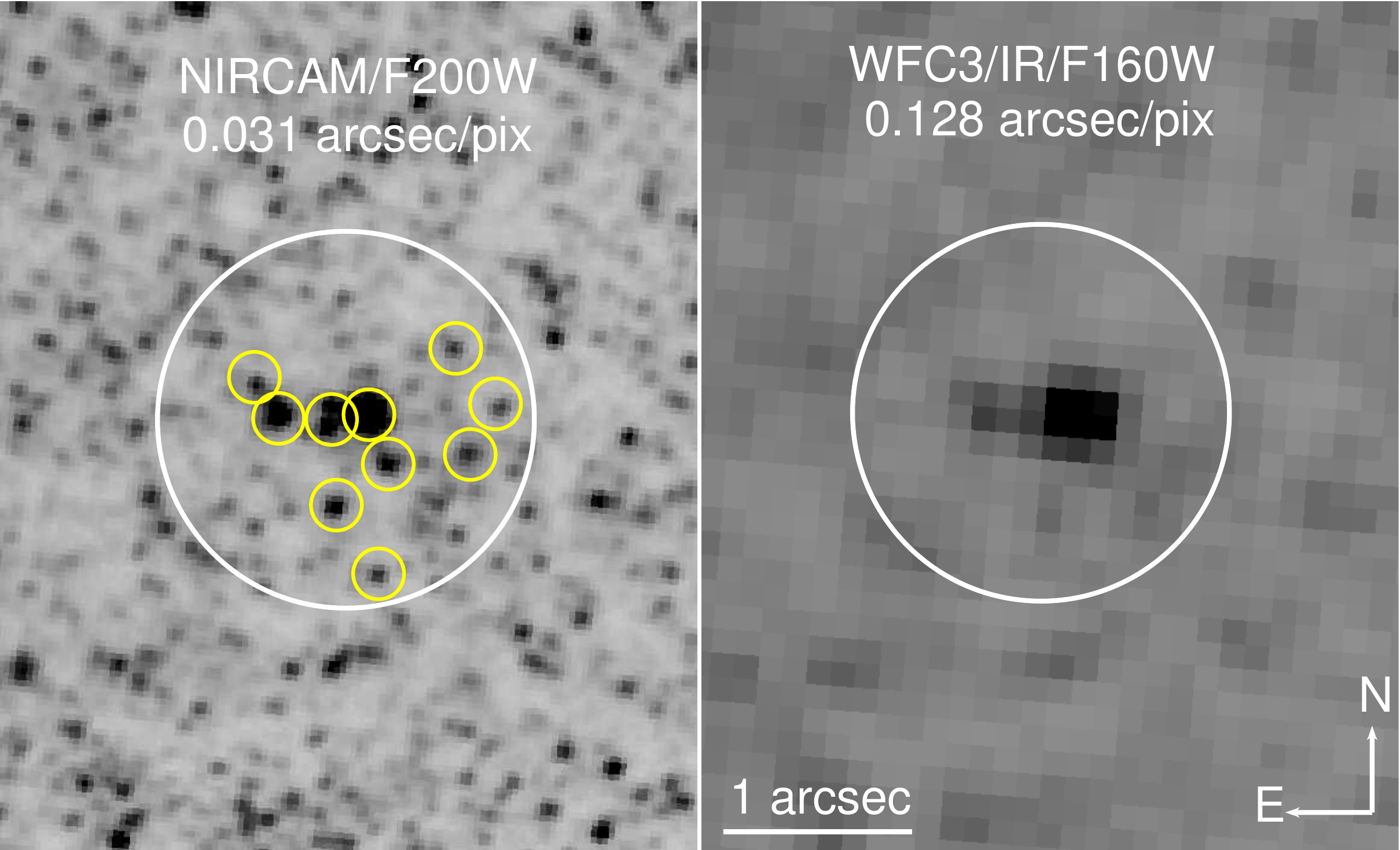}
\includegraphics[width=\columnwidth]{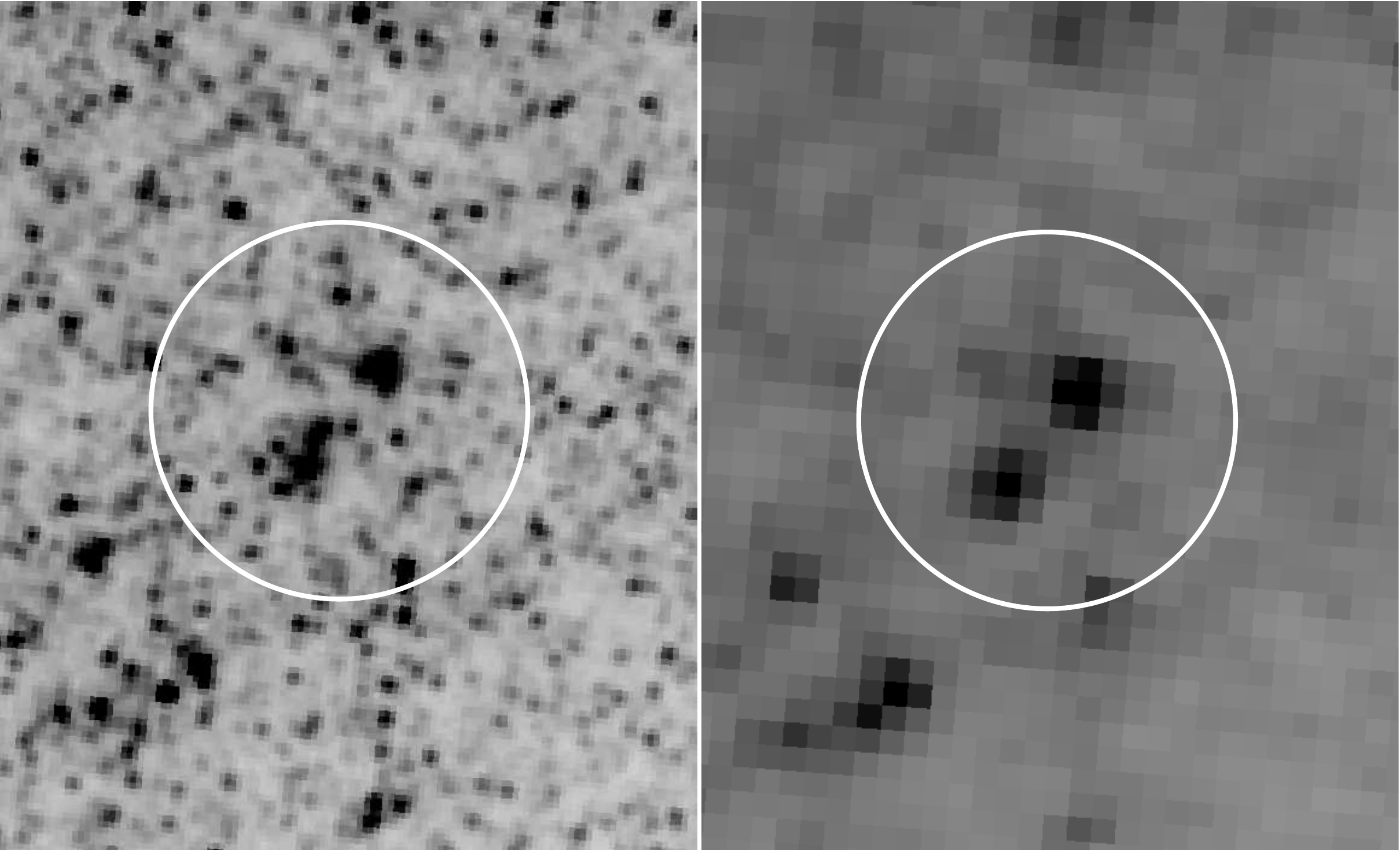}
\caption{Top two panels: Comparison of {\it JWST} {\it MIRI} F2100Wimage (left) (0.111 arcsec/pixel) with {\it NIRcam} F200W image (0.031 arcsec/pixel) (right). Bottom two panels: Comparison of {\it JWST} {\it MIRI} F2100W image (left) with {\it NIRcam} F300M image (0.063 arcsec/pixel) (right). The sources were chosen randomly.}
\label{F:compare}
\end{center}
\end{figure}

Since the ULXs are located in crowded regions (in terms of point sources), \textit{JWST/MIRI} instrument does not have sufficient angular resolution to resolve counterparts at this distance of galaxy NGC 1566. Therefore, Mid-IR counterparts were not identified in the \textit{JWST/MIRI} images. Moreover, although, there is almost no difference between the spatial resolution of the \textit{JWST/NIRCam} F200W (0.03 arcsec/pixel) and \textit{HST/WFC3} (0.04 arcsec/pixel), as seen in Figure \ref{F:ulx4}, two NIR sources are detected in the F200W image while a unique source is observed at the same location in the HST/WFC3 image. Furthermore, as seen in Figure \ref{F:compare}, randomly selected sources that can be resolved in the \textit{JWST/NIRcam} images are observed as bright point-like sources in the \textit{JWST/MIRI} images (0.11 arcsec/pixel). Due to the fact that ULXs are often found in crowded fields, such as spiral arms, it is quite possible that past IR studies of ULXs especially in distant galaxies might be identified blended sources as counterpart(s). Therefore, the previously reported IR properties of ULXs should be re-examined with \textit{JWST} observations.\\

\subsection{Properties of Counterparts}

\textit{ULX-1}: The SED (optical to NIR) for the counterpart of ULX-1 is fitted with a \textit{blackbody} model with \textit{T} $\sim$ 1500 K. This indicates that the emission of ULX-1 could be due to the accretion disk, the donor star, or the circumbinary disk/dust. On the other hand, the {\it blackbody} temperature (1500 K), which represents the NIR excess, is not hot enough to come from any donor star (e.g., \citealp{2005ApJ...628..973L}) so, it is more likely that this temperature comes from the circumbinary disk or from warm dust disturbed by X-rays \citep{2019ApJ...878...71L}. Since the emissions at optical wavelengths are generally the same size as local dust particles, it is easily scattered by the dust while longer-wavelength emission passes through unobstructed. Therefore, the optical counterpart of ULX-1 might be observed as very faint ($\sim$ 26 mag) in the {\it HST/WFC3} F555W image. Since the observed emission does not come from the outer part of the accretion disk (reprocessing) and/or the donor star, the multi-wavelength properties of ULX-1 are still unclear.\\

\textit{ULX-2}: The SED for the counterpart of 2-b is well-fitted with a \textit{power-law} model with {$\alpha$} = 1.9. This indicates that the emission from of ULX-2 could be due to the accretion disk. The \textit{power-law} does not cut off at optical wavelengths and extends into the NIR. In the case of the accretion disk, such an emission indicates the presence of a quite large ($\geq$ 10$^{13}$ cm) disk than expected for ULXs \citep{2011ApJ...737...81T,2012ApJ...745..123G,2014MNRAS.444.2415S} therefore, the source 2-b may not be the possible donor star of ULX-2. On the other hand, another counterpart of ULX-2 (2-a ) is not detected at the 3-$\sigma$ detection threshold in the \textit{HST/WFC3} images hence, for \textit{HST/WFC3} F555W and F814W images, the 3-$\sigma$ upper limit magnitude values of 2-a were derived as 26.8 and 28 , which are close to the detection limit of {\it HST/WFC3} instrument. The SED of 2-a was represented by the \textit{blackbody} model with \textit{T}=700 K. As seen in the top panel of Figure \ref{F:SED2}, the only NIR excess is represented by the \textit{blackbody} model. This thermal emission is due to the presence of circumbinary disk/dust similar to the case of ULX-1, but colder.

\begin{figure}
\begin{center}
\includegraphics[width=\columnwidth]{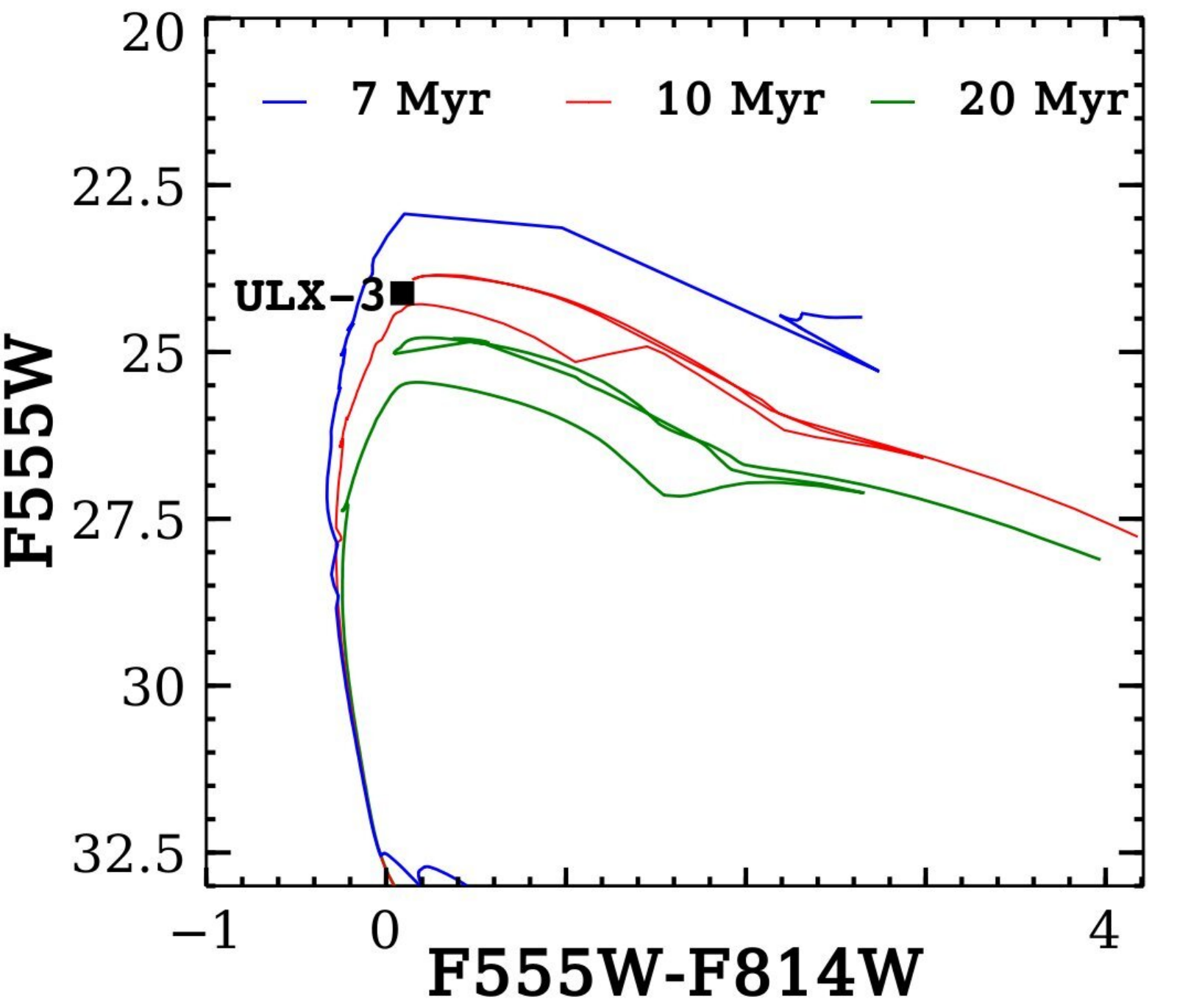}
\caption{CMD {\it HST/WFC3} F555W vs F555W - F814W for counterpart of ULX-3. The isochrones were corrected for the extinction of A$_{V}$ = 0.04 mag.}
\label{F:cmd1}
\end{center}
\end{figure}

\textit{ULX-3}: The optical counterpart of ULX-3 is bright in the optical bands and there are no \textit{JWST} observations to investigate its IR counterpart(s). As seen in the reddening corrected SEDs of ULX-3, Figure \ref{F:SED1}, optical emission was represented by the \textit{blackbody} model with a temperature of 18000 K.According to \cite{1981Ap&SS..80..353S} the measured temperature and absolute magnitude (M$_{V} \sim$ -7) of the possible donor star indicate that spectral classification would be an B-type supergiant. In addition, its colors (U-B and V-I) are also consistent with the B-type supergiants in the Large Magellanic Cloud \citep{2009AJ....138.1003B}. Moreover, to estimate age of the counterparts, color-magnitude diagrams (CMDs) were plotted for the counterpart. The solar metallicity of 0.02 and A$_{V}$ $=$0.04 mag \citep{2022ApJ...925..101T} were used to obtain the corresponding Vega mag PARSEC \citep{2012MNRAS.427..127B} isochrones for{\it HST/WFC3} wide filters. The distance modulus was calculated as 32.22 using the adopted distance of $\sim$ 17.7 Mpc. CMD, F555W vs F555W$-$F814W was plotted for the counterpart, and the point sources around ULX-3 (see Figure \ref{F:cmd1}).According to CMD, the range of age of ULX-3 was estimated as $\sim$ 7-10 Myr. If indeed the isochrones represent this source, in other words, if the radiation comes from possible donor, the mass range of the optical counterpart should be limited to 10-20 M$\odot$. In typical Galactic BHB, the X-ray flux from the inner source is reprocessed within the outer parts of the accretion disk, and this can observed as a dominant effect on the UV and optical spectrum. The surface area of the emitting regions of ULX-3 was derived as 8 $\times$ 10$^{22}$ {\it cm$^{2}$} or a radius of 3 $\times$ 10$^{11}$ {\it cm} using temperatures of 18000 K. This radius is expected for many ULXs \citep{2011ApJ...737...81T,2012ApJ...745..123G,2012ApJ...750..110T}, hence this study rule out the possibility of reprocessing. In this case, the optical emission observed for ULX-3 is most likely due to the donor star.

\section{CONCLUSIONS AND SUMMARY} \label{sec:5}

The timing and spectral properties of the four ULXs in the NGC 1566 analyzing all the available archival {\it Swift/XRT} PC mode and \textit{Chandra} observations. Moreover, the optical and NIR counterparts of ULXs were identified by precise astrometric calculations by using {\it HST} \textit{HST/WFC3} and {\it JWST/NIRCam} images. To constrain the optical and NIR emissions of the counterparts, SEDs were plotted using all available multi-wavelength observations. Moreover, CMDs were constructed to estimate the age and type of donor star. The main findings from this study are summarized as follows:\\

\begin{itemize}

\item new transient ULX candidate (ULX-4) with a peak luminosity of$\sim$10$^{40}$ erg s$^{-1}$ was identified.\\

\item For ULX-1, ULX-2, and ULX-3, no strong evidence of long-term variability is found, while ULX-4 shows variability of more than an order of magnitude in the long-term X-ray light curve. In addition, a hint of short-term (3 ks) variability was found in the {\it Chandra} data for the only ULX-1.\\

\item The transition track of ULX-3 is most likely the spectral state transition seen in Galactic X-ray binaries.\\

\item The time-averaged spectrum of the source ULX-2 was adequately well-fitted by {\it power-law} model with $\Gamma$ $\sim$ 1.7,and L$_{X} \sim$ 10$^{40}$ \textit{erg} s$^{-1}$.\\

\item The time-averaged spectrum of the source ULX-3 was well-fitted by the {\it power-law+blackbody} model with $\Gamma$ $\sim$ 2 and \textit{kT} =0.12 keV, and L$_{X} \sim$ 4 $\times$ 10$^{39}$ \textit{erg} s$^{-1}$.\\

\item The time-averaged spectrum of the source ULX-4 was well-fitted by the {\it power-law+diskbb} model with $\Gamma$ $\sim$ 1.6 and \textit{Tin} =0.26 keV, and L$_{X} \sim$ 10$^{40}$ \textit{erg} s$^{-1}$.\\

\item A unique optical and NIR counterparts of ULX-1 were identified within the astrometric error radius while two NIR and optical counterparts, 2-a and 2-b, were identified for ULX-2. At least seven NIR sources were identified for ULX-4 within the error radius. In the case of ULX-3, a unique optical counterpart was identified.\\

\item The SEDs for the NIR counterparts ULX-1 and ULX-2-a were adequately well-fitted by a single {\it blackbody} model with temperatures of 1500 K and 700 K, respectively. These indicate that observed NIR emission comes from the circumbinary disk or from warm dust. \\

\item The SED of ULX-3 is represented by a single {\it blackbody} temperature of 18000 K. Both the absolute magnitude and temperature indicate that the observed optical emission comes from a B-type supergiant donor.\\

\end{itemize}

\section*{Acknowledgements}
\noindent
I would like to thank the anonymous referee for helpful suggestions that greatly improved this paper. This paper was supported by the Scientific and Technological Research Council of Turkey (TÜBİTAK) through project number 122C183. I would like to thank A. Akyuz for her valuable suggestions.

\section*{Data Availability}

The scientific results reported in this article are based on archival observations made by the James Webb Space Telescope and Hubble Space Telescope, and obtained from the data archive at the Space Telescope Science Institute\footnote{https://mast.stsci.edu/portal/Mashup/Clients/Mast/Portal.html}. This work has also made use of observations made with the {\it Chandra}\footnote{https://cda.harvard.edu/chaser/} and {\it Swift/XRT}\footnote{https://swift.gsfc.nasa.gov/about\_swift/xrt\_desc.html} X-ray Observatories.

\bibliographystyle{mnras}
\bibliography{ngc1566} 

\bsp	
\label{lastpage}
\end{document}